\documentclass[structabstract]{aa}
\usepackage{graphicx}
\usepackage{txfonts}
\usepackage{natbib}
\usepackage{array}
\usepackage{color}
\bibpunct{(}{)}{;}{a}{}{,}

\begin{document}

   \title{On the shape of the spectrum of cosmic rays \\accelerated inside superbubbles}
   \titlerunning{The spectrum of cosmic rays accelerated inside superbubbles}

   \author{Gilles Ferrand \inst{1} \and Alexandre Marcowith\inst{2}}
   \authorrunning{Ferrand and Marcowith}

   \institute{Laboratoire Astrophysique Interactions Multi-{\'e}chelles (AIM), CEA/Irfu, CNRS/INSU, Universit{\'e} Paris~VII \\
   		L'Orme des Merisiers, b{\^a}t.~709, CEA Saclay, 91191~Gif-sur-Yvette Cedex, France\\
                 \email{g.ferrand@cea.fr}
         \and
             Laboratoire de Physique Th{\'e}orique et Astroparticules (LPTA), CNRS/IN2P3, Universit{\'e} Montpellier~II \\ 
             Place Eug{\`e}ne Bataillon, 34095 Montpellier C{\'e}dex, France\\
             \email{alexandre.marcowith@lpta.in2p3.fr}
             }

   \date{2009/12/10}
 
  \abstract
   {Supernova remnants are believed to be a major source of energetic particles (\emph{cosmic rays}) on the Galactic scale. Since their progenitors, namely the most massive stars, are commonly found clustered in \emph{OB~associations}, one has to consider the possibility of collective effects in the acceleration process.}
   {We investigate the shape of the spectrum of high-energy protons produced inside the \emph{superbubbles} blown around clusters of massive stars.}
   {We embed simple semi-analytical models of particle acceleration and transport inside Monte Carlo simulations of OB~associations timelines. We consider regular acceleration (Fermi~1 process) at the shock front of supernova remnants, as well as stochastic reacceleration (Fermi~2 process) and escape (controlled by magnetic turbulence) occurring between the shocks. In this first attempt, we limit ourselves to linear acceleration by strong shocks and neglect proton energy losses.}
   {We observe that particle spectra, although highly variable, have a distinctive shape because of the competition between acceleration and escape: they are harder at the lowest energies (index $s<4$) and softer at the highest energies ($s>4$). The momentum at which this spectral break occurs depends on the various bubble parameters, but all their effects can be summarized by a single dimensionless parameter, which we evaluate for a selection of massive star regions in the Galaxy and the LMC.}
   {The behaviour of a superbubble in terms of particle acceleration critically depends on the magnetic turbulence: if~B is low then the superbubble is simply the host of a collection of individual supernovae shocks, but if~B is high enough (and the turbulence index is not too high), then the superbubble acts as a global accelerator, producing distinctive spectra, that are potentially very hard over a wide range of energies, which has important implications on the high-energy emission from these objects.}

   \keywords{acceleration of particles -- shock waves -- turbulence -- cosmic rays -- supernova remnants }

   \maketitle


\section{Introduction}
\label{sec:introduction}

Superbubbles are hot and tenuous large structures that are formed around OB~associations 
by the powerful winds and the explosions of massive stars \citep{Higdon2005a}.
They are the major hosts of supernovae in the Galaxy, and thus major candidates 
for the production of energetic particles (e.g., \citealt{Montmerle1979a}, \citealt{Bykov2001b},
\citealt{Butt2009a}, and references therein). 
Supernovae are indeed believed to be the main contributors of Galactic cosmic rays 
(along with pulsars and micro-quasars), by means of the \emph{diffusive shock acceleration} 
process (a 1st-order, regular Fermi process) occurring at the remnant's blast wave
as it goes through the interstellar medium \citep{Drury1983a,Malkov2001c}.

Supernovae in superbubbles are correlated in space and time, hence the need to investigate 
acceleration by multiple shocks \citep{Parizot2004a}. 
\citet{Klepach2000a} developed a semi-analytical model of test-particle acceleration 
by multiple spherical shocks 
(either wind termination shocks, or supernova shocks plus wind external shocks),
based on the limiting assumption of small shocks filling factors.
\citet{Ferrand2008a} performed direct numerical simulations of repeated acceleration 
by successive planar shocks in the non-linear regime (that is, taking into account the back-reaction
of energetic particles on the shocks). However, to ascertain the particle spectrum 
produced inside the superbubble as a whole, one must also consider important
physics occurring \emph{between} the shocks. Since the bubble interior is probably magnetized 
and turbulent, we need to evaluate gains and losses caused by the acceleration by waves 
(a 2nd-order, stochastic Fermi process) and escape from the bubble.

In this study, we combine the effects of regular acceleration (occurring quite discreetly, at shock fronts) 
and stochastic acceleration and escape (occurring continuously, between shocks),
to determine the typical spectra that we can expect inside superbubbles over the lifetime
of an OB~cluster. We choose to treat regular acceleration as simply as we can, 
and concentrate on modeling the relevant scales of stochastic acceleration and escape 
inside superbubbles. We present our model in Sect.~\ref{sec:model}, give our general results 
in Sect.~\ref{sec:results}, and present specific applications in Sect.~\ref{sec:application}. 
Finally we discuss the limitations of our approach in Sect.~\ref{sec:limitations} and
provide our conclusions in Sect.~\ref{sec:conclusions}.


\section{Model}
\label{sec:model}

Our model is based on Monte Carlo simulations of the activity of a cluster of massive stars, 
in which we embed simple semi-analytical models of (re-)acceleration and escape
(described by means of their Green functions).
To evaluate the average properties of a cluster of~$N_{\star}$ stars,
we perform random samplings of the initial mass function (Sect.~\ref{sec:OB}). 
For a given cluster, time is sampled in intervals $\mathrm{d}t=10\:000\:\mathrm{yr}$, 
which is short enough to ensure that at most one supernova occurs during that period,
but by chance for large clusters, and which is long enough to consider that
regular acceleration at a shock front has shaped the spectrum of particles
-- acceleration is thought to take place mostly at early stages of supernova remnant evolution,
and in a superbubble the Sedov phase begins after a few thousands of years \citep{Parizot2004a}.
Here we do not try to investigate the exact extent of the spectrum
of accelerated particles: we set the lowest momentum (injection momentum)
to be $p_{\mathrm{min}}=10^{-2}\: m_{p}c$ (which is the typical thermal
momentum downstream of a supernova shock) and set the highest momentum
(escape momentum) to be $p_{\mathrm{max}}=10^{6}\: m_{p}c\simeq10^{15}\:\mathrm{eV}$
(which corresponds to the ``knee" break in the spectrum of cosmic rays as
observed on the Earth). We note that the theoretical acceleration time
from $p_{\mathrm{min}}$ to $p_{\mathrm{max}}$ (in the linear regime,
without escape) is roughly 8~000~yr (assuming Bohm diffusion with
$B=10\:\mu G$), which is again consistent with our choice of $\mathrm{d}t$.
This corresponds to 8 decades in~$p$, at a resolution of a
few tens of bins per decade (according to Sect.~\ref{sec:Fermi1_multiple}). 

The procedure is then as follows: for each time bin in the life of the cluster,
either~(1) a supernova occurs, and the distribution of particles evolves according 
to the diffusive shock acceleration process, as explained in Sect.~\ref{sec:Fermi1};
or~(2) no supernova occurs, and the distribution evolves taking into account 
acceleration and escape controlled by magnetic turbulence, 
as explained in Sect.~\ref{sec:Fermi2}.
This process is repeated for many random clusters of the same size,
until some average trend emerges regarding the shape of spectra 
(note that average spectra are not monitored for each bin $\mathrm{d}t$
but in larger steps of 1~Myr).

In the following, we describe our modeling of massive stars, supernovae shocks, 
and magnetic turbulence.


\subsection{OB clusters: random samplings of supernovae}
\label{sec:OB}

We are interested in massive stars that die by core-collapse, 
producing type~Ib,~Ic~or~II supernovae, that is of mass greater than 
$m_{\mathrm{min}}=8\: m_{\odot}$, and up to say $m_{\mathrm{max}}=120\: m_{\odot}$.
These are stars of spectral type~O ($>20\: m_{\odot}$)
and include stars of spectral type~B ($4-20\: m_{\odot}$).
Most massive stars spend all their life within the cluster in which they were born, 
forming OB~associations. To describe the evolution of such a cluster, 
one needs to know the distribution of star masses and lifetimes.

\begin{figure}[!t]
\resizebox{\hsize}{!}{\includegraphics{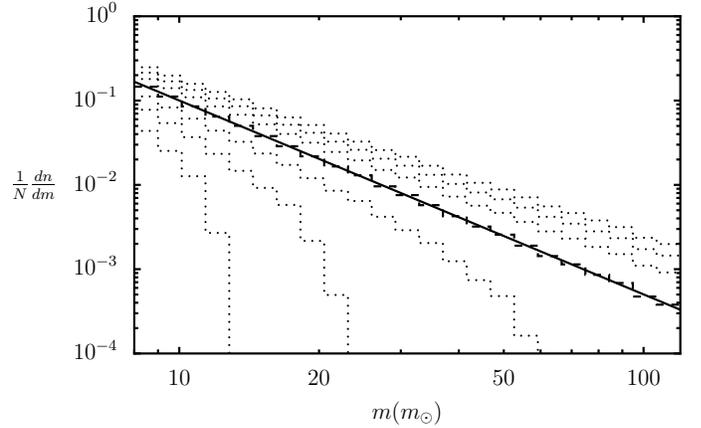}}
\caption{Distribution of massive stars masses: the initial mass function.
For each cluster $N_{\star}=100$~stars are randomly chosen in the IMF~(\ref{eq:IMF}). 
The dashed curve represents the experimental histogram of masses
after $N_{\mathrm{OB}}=1000$~samples (with resolution $\mathrm{d}\log m=0.05$). 
The dotted curves show 1-, 2-, 3-sigma standard deviations
over the clusters set. The solid curve is the theoretical IMF.}
\label{fig:IMF}
\end{figure}

The initial mass function (IMF) $\xi$ is defined so that the number of stars in the mass interval
$m$ to $m+\mathrm{d}m$ is $\mathrm{d}n=\xi\left(m\right)\times\mathrm{d}m$,
so that the number of stars of masses between $m_{\mathrm{min}}$
and $m_{\mathrm{max}}$ is
$N_{\star}=\int_{m_{\mathrm{min}}}^{m_{\mathrm{max}}}\xi\left(m\right)\:\mathrm{d}m\:.$
Observations show that $\xi$ can be expressed as a power law (\citealt{Salpeter1955a})
\begin{equation}
\label{eq:IMF}
\xi\left(m\right) \propto m^{\alpha}\:,
\end{equation}
with an index of $\alpha=2.30$ for massive stars (\citealt{Kroupa2002a}).
This function is shown in Fig.~\ref{fig:IMF}.

\begin{figure}[!t]
\resizebox{\hsize}{!}{\includegraphics{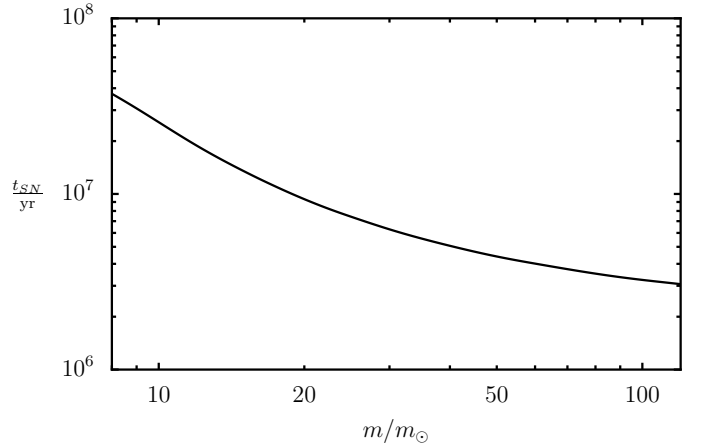}}
\caption{Distribution of massive stars lifetimes (data from \citet{Limongi2006a}).}
\label{fig:lifetimes}
\end{figure}

Stars lifetimes can be computed from stellar evolution models, 
and here we use data from \citet{Limongi2006a}, which is plotted in Fig.~\ref{fig:lifetimes}.
The more massive they are, the faster stars burn their material. 
A star at the threshold $m_{\mathrm{min}}=8\: m_{\odot}$ has a lifetime of 
$t_{\mathrm{SN},\mathrm{max}}\simeq37\:\mathrm{Myr}$, 
which is also the total lifetime of the cluster;
a star of $m_{\mathrm{max}}=120\: m_{\odot}$ lives only 
$t_{\mathrm{SN},\mathrm{min}}\simeq3\:\mathrm{Myr}$.
Regarding supernovae, the active lifetime of the cluster is thus
\begin{equation}
\label{eq:OB_lifetime}
\Delta t_{\mathrm{OB}}^{\star}=t_{\mathrm{SN}}\left(m_{\mathrm{min}}\right)-t_{\mathrm{SN}}\left(m_{\mathrm{max}}\right)\simeq34\:\mathrm{Myr}\:.
\end{equation}

\subsection{Supernovae shocks: regular acceleration}
\label{sec:Fermi1}

\subsubsection{Green function}
\label{sec:Fermi1_Green}

To keep things as simple as possible, we limit ourselves here to the test-particle approach
(non-linear calculations will be presented elsewhere).
In the linear regime, we know the Green function $G_1$ that links the distributions\footnote{
The distribution function $f(p)$ is defined so that the particles number density is
$n=\int_{p}f\left(p\right)\:4\pi p^{2}\:\mathrm{d}p$,
where $p$ is the momentum.
} of particles
downstream and upstream of a single shock according to
\begin{equation}
f_{\mathrm{down}}\left(p\right)=\int_{0}^{\infty}G_{1}\left(p,p_{0}\right)\: f_{\mathrm{up}}\left(p_{0}\right)\:\mathrm{d}p_{0}\:;
\label{eq:G1_def}
\end{equation}
it reads
\begin{equation}
G_{1}\left(p,p_{0}\right)=\frac{s_{1}}{p_{0}}\left(\frac{p}{p_{0}}\right)^{-s_{1}}H\left(p-p_{0}\right)\label{eq:G1_expr}
\end{equation}
where $H$ is the Heaviside function, and 
\begin{equation}
\label{eq:s1}
s_{1} = \frac{3r}{r-1}\:,
\end{equation}
where $r$ is the compression ratio of the shock.

\subsubsection{Adiabatic decompression}
\label{sec:Fermi1_multiple}

Around an OB~association, particles produced by a supernova shock 
might be reaccelerated by the shocks of subsequent supernova
before they escape the superbubble. 
The effect of repeated acceleration is basically to harden the spectra
(\citealt{Achterberg1990a}, \citealt{Melrose1993a}).

When dealing with multiple shocks, it is mandatory
to account for adiabatic decompression between the shocks: 
the momenta of energetic particles bound to the fluid will decrease by a factor $R=r^{1/3}$ 
when the fluid density decreases by a factor $r$. 
To resolve decompression properly, the numerical momentum resolution 
$\mathrm{d}\log p$ has to be significantly smaller than the induced momentum shift (\citealt{Ferrand2008a}).


\subsection{Magnetic turbulence: stochastic acceleration and escape}
\label{sec:Fermi2}

Particles accelerated by supernova shocks, although energetic,
might remain for a while inside the superbubble because of magnetic turbulence
that scatters them (they perform a random walk until they escape).
Because of this turbulence, particles will also experience stochastic reacceleration 
during their stay in the bubble. We present here a deliberately simple model of transport, 
to obtain the relevant functional dependences and order of magnitudes of the diffusion coefficients.
The turbulent magnetic field $\delta B$ is represented by its power spectrum $W(k)$,
defined so that 
$\delta B^{2} \propto \int_{k_{\mathrm{min}}}^{k_{\mathrm{max}}}W(k)\:\mathrm{d}k$,
where $k=2\pi/\lambda$, $\lambda$ is the turbulence scale,
and $k_{\mathrm{min}}$ (respectively $k_{\mathrm{max}}$) corresponds to waves interacting 
with the particles of highest (respectively lowest) energy. 
This spectrum is usually taken to be a power law of index $q$
\begin{equation}
\label{eq:W(k)}
W(k)\propto k^{-q}\:,
\end{equation}
normalised by the turbulence level 
\begin{equation}
\label{eta}
\eta_{T} = \frac{{\left\langle \delta B^{2}\right\rangle }}{{B^{2}+\left\langle \delta B^{2}\right\rangle}} .
\end{equation}

\subsubsection{Diffusion scales}
\label{sec:Fermi2_scales}

If the turbulence follows Eq.~(\ref{eq:W(k)}), then the space diffusion coefficient is given by
\begin{equation}
D_{x}\left(p\right) = D_{x}^{\star}\times\left(\frac{p}{m_{p}c}\right)^{2-q}\:,
\label{eq:Dx}
\end{equation}
where we assume that the turbulence spectrum extends sufficiently for this description to remain correct at the lowest particle energies.
Using results from \citet{Casse2002a} obtained for isotropic turbulence, one can assume that
\begin{equation}
D_{x}^{\star}\propto\eta_{T}^{-1}\: B^{q-2}\:\lambda_{\mathrm{max}}^{q-1}\:.
\label{eq:Dx*}
\end{equation}
For standard turbulence indices, we obtain
\begin{equation}
\label{eq:Dx_ref}
\frac{D_{x}\left(p\right)}{10^{26}\:\mathrm{cm^{2}.s^{-1}}} \simeq
\left\{
\begin{array}{ll}
\frac{12,3}{\eta_{T}}\left(\frac{B}{10\:\mathrm{\mu G}}\right)^{-\frac{1}{3}}\left(\frac{\lambda_{\mathrm{max}}}{10\:\mathrm{pc}}\right)^{\frac{2}{3}}\left(\frac{p}{m_{p}c}\right)^{\frac{1}{3}} & q=5/3\\
\frac{0,8}{\eta_{T}}\left(\frac{B}{10\:\mathrm{\mu G}}\right)^{-\frac{1}{2}}\left(\frac{\lambda_{\mathrm{max}}}{10\:\mathrm{pc}}\right)^{\frac{1}{2}}\left(\frac{p}{m_{p}c}\right)^{\frac{1}{2}} & q=3/2
\end{array}
\right.\:.
\end{equation}
Particles diffuse over a typical length scale of $x_{\mathrm{diff}}=\sqrt{6\: D_{x}\: t}$.
They are confined within the acceleration region of size $x_{\mathrm{acc}}$
as long as $x_{\mathrm{diff}}\left(t\right)<x_{\mathrm{acc}}$, 
hence a typical escape time is $t_{\mathrm{esc}}={x_{\mathrm{acc}}^{2}}/{6\: D_{x}}$,
that is, using Eq.~(\ref{eq:Dx})
\begin{equation}
t_{\mathrm{esc}}\left(p\right) = t_{\mathrm{esc}}^{\star}\times\left(\frac{p}{m_{p}c}\right)^{q-2}\:,
\label{eq:Tesc}
\end{equation}
where
\begin{equation}
t_{\mathrm{esc}}^{\star}\propto\eta_{T}\: B^{2-q}\:\lambda_{\mathrm{max}}^{1-q}\: x_{\mathrm{acc}}^{2}\:.
\label{eq:Tesc*}
\end{equation}
For standard turbulence indices, we obtain
\begin{equation}
\label{eq:Tesc_ref}
\frac{t_{\mathrm{esc}}\left(p\right)}{10^{13}\:\mathrm{s}} \simeq
\left\{
\begin{array}{ll}
\frac{\eta_{T}}{5,0}\left(\frac{B}{10\:\mathrm{\mu G}}\right)^{\frac{1}{3}}\left(\frac{\lambda_{\mathrm{max}}}{10\:\mathrm{pc}}\right)^{-\frac{2}{3}}\left(\frac{x_{\mathrm{acc}}}{40\:\mathrm{pc}}\right)^{2}\left(\frac{p}{m_{p}c}\right)^{-\frac{1}{3}} & q=5/3\\
\frac{\eta_{T}}{0,3}\left(\frac{B}{10\:\mathrm{\mu G}}\right)^{\frac{1}{2}}\left(\frac{\lambda_{\mathrm{max}}}{10\:\mathrm{pc}}\right)^{-\frac{1}{2}}\left(\frac{x_{\mathrm{acc}}}{40\:\mathrm{pc}}\right)^{2}\left(\frac{p}{m_{p}c}\right)^{-\frac{1}{2}} & q=3/2
\end{array}
\right.\:.
\end{equation}

Interaction with waves also leads to a diffusion in momentum. 
Using results from quasi-linear theory, we can express the diffusion coefficient as
\begin{equation}
D_{p}\left(p\right) = D_{p}^{\star}\times(m_{p}c)^{2}\times\left(\frac{p}{m_{p}c}\right)^{q}\:,
\label{eq:Dp}
\end{equation}
where
\begin{equation}
D_{p}^{\star}\propto\eta_{T}\: B^{4-q}\:\lambda_{\mathrm{max}}^{1-q}\: n^{-1}\:.
\label{eq:Dp*}
\end{equation}
and $n$ is the number density (which determines the Alfv{\'e}n velocity together with~$B$).
For standard turbulence indices, we obtain
\[
\label{eq:Dp_ref}
\frac{D_{p}\left(p\right)}{10^{-38}\:\mathrm{g^2.cm^2.s^{-3}}} \simeq
\]
\begin{equation}
\left\{
\begin{array}{ll}
\frac{\eta_{T}}{20}\left(\frac{B}{10\:\mathrm{\mu G}}\right)^{\frac{7}{3}}\left(\frac{\lambda_{\mathrm{max}}}{10\:\mathrm{pc}}\right)^{-\frac{2}{3}}\left(\frac{n}{10^{-2}\:\mathrm{cm^{-3}}}\right)^{-1}\left(\frac{p}{m_{p}c}\right)^{\frac{5}{3}} & q=5/3\\
\frac{\eta_{T}}{1,4}\left(\frac{B}{10\:\mathrm{\mu G}}\right)^{\frac{5}{2}}\left(\frac{\lambda_{\mathrm{max}}}{10\:\mathrm{pc}}\right)^{-\frac{1}{2}}\left(\frac{n}{10^{-2}\:\mathrm{cm^{-3}}}\right)^{-1}\left(\frac{p}{m_{p}c}\right)^{\frac{3}{2}} & q=3/2
\end{array}
\right.\:.
\end{equation}

\subsubsection{Green function}
\label{sec:Fermi2_Green}

\citet{Becker2006a} presented the first analytical expression of 
the Green function $G_{2}$ for both stochastic acceleration and escape
that is valid for any turbulence index $q\in]0,2[$.
It is defined so that, for impulsive injection of distribution $f_{\mathrm{init}}$,
the distribution after time~$t$ is
\begin{equation}
f_{\mathrm{end}}\left(p,t\right)=\int_{0}^{\infty}G_{2}\left(p,p_{0},t\right)\: f_{\mathrm{init}}\left(p_{0}\right)\:\mathrm{d}p_{0}\:.
\label{eq:G2_def}
\end{equation}
Neglecting losses, it can be expressed as
\begin{eqnarray}
G_{2}\left(p,p_{0},t\right) & = & \frac{2-q}{p_{0}}\:\sqrt{\frac{p}{p_{0}}}\:\frac{\sqrt{zz_{0}\xi}}{1-\xi}\:
\label{eq:G2_expr} \\
 & \times & \exp\left(-\frac{\left(z+z_{0}\right)\left(1+\xi\right)}{2\left(1-\xi\right)}\right)\:\mathrm{I}\left(\frac{1+q}{2-q},\frac{2\sqrt{zz_{0}\xi}}{1-\xi}\right)\nonumber\:, \\
z\left(p\right) & = &{ 2\: p^{2-q} } / \left( \left(2-q\right) \sqrt{D_{\mathrm{p}}^{\star}t_{\mathrm{esc}}^{\star}} \right)
 \nonumber\:, \\
\xi\left(t\right) & = & \exp\left(2\left(q-2\right)\: 
{ D_{\mathrm{p}}^{\star}t\: } / { \sqrt{D_{\mathrm{p}}^{\star}t_{\mathrm{esc}}^{\star}} }
\right)\nonumber\:,
\end{eqnarray}
where $\mathrm{I}\left(o,x\right)$ is the modified Bessel function of the first kind,
and we recall that $D_{\mathrm{p}}^{\star}$ and $t_{\mathrm{esc}}^{\star}$ 
are defined by Eqs.~(\ref{eq:Dp*}) and~(\ref{eq:Tesc*}) respectively.

$G_{2}$ represents the distribution of particles remaining inside the bubble.
One can also evaluate the rate of particles escaping the bubble by
dividing $G_{2}$ by the escape time given by Eq.~(\ref{eq:Tesc}):
\begin{equation}
\dot{G}_{2,\mathrm{esc}}\left(p,p_{0},t\right)=\frac{G_{2}\left(p,p_{0},t\right)}{t_{\mathrm{esc}}\left(p\right)}=\frac{p^{2-q}\: G_{2}\left(p,p_{0},t\right)}{t_{\mathrm{esc}}^{\star}}\:.\label{eq:G2_esc}
\end{equation}


\section{Results}
\label{sec:results}


\subsection{Distribution of supernovae shocks}
\label{sec:res_shocks}

Before presenting the spectra of particles, we briefly discuss the temporal distribution of shocks 
during the life of the cluster, because this controls the possibility of repeated acceleration.

\subsubsection{Rate of supernovae}
\label{sec:SN_rate}

\begin{figure}[!t]
\resizebox{\hsize}{!}{\includegraphics{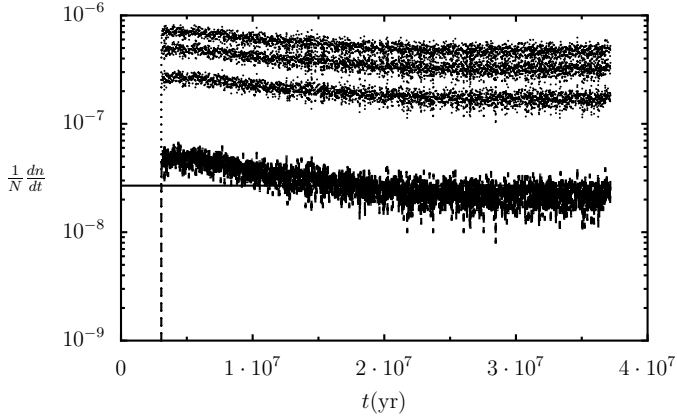}}
\caption{Mean supernovae rate as a function of time.
For each cluster, $N_{\star}=100$~stars are randomly chosen in the IMF.
The central curve represents the experimental mean rate of supernovae
after $N_{\mathrm{OB}}=1000$~samples (with resolution $\mathrm{d}t=10^4$ years). 
The top curves show 1, 2 and 3 standard deviations over the clusters set.
The solid curve is the theoretical mean rate of supernovae over the cluster's active lifetime~(\ref{eq:OB_lifetime}), 
i.e. $N_{\star}/\left(t_{\mathrm{SN,max}}-t_{\mathrm{SN,min}}\right)$.}
\label{fig:SN_rate}
\end{figure}

As an illustration of our Monte Carlo procedure, if we count the number
of supernovae in each time bin $\left[t,t+\mathrm{d}t\right]$,
we can estimate the mean supernovae rate. 
The result is shown in Fig.~\ref{fig:SN_rate}.
In agreement with the ``instantaneous burst" model of \citet{Cervino2000a}, 
we observe that the distributions of masses and lifetimes
combine in such a way that, but for a peak at the beginning, the rate of supernovae is fairly constant
during the cluster's life, and can be expressed to a first approximation by 
\begin{equation}
\label{eq:SN_rate}
\frac{{\mathrm{d}n_{\mathrm{SN}}}}{{\mathrm{d}t}} \simeq \frac{N_{\star}}{\Delta t_{\mathrm{OB}}^{\star}}
\simeq  N_{\star}\times3.10^{-8}\ \mathrm{yr^{-1}}\:,
\end{equation}
where we recall that $\Delta t_{\mathrm{OB}}^{\star}$ is the active
lifetime of the cluster, given by Eq.~(\ref{eq:OB_lifetime}).

\subsubsection{Typical time between shocks}
\label{sec:delta_shocks}

\begin{figure}[!t]
\resizebox{\hsize}{!}{\includegraphics{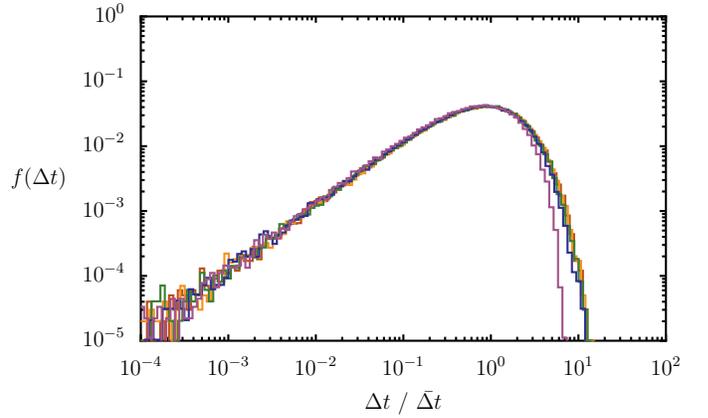}}
\caption{Distribution of the interval between two successive shocks 
(normalised to the average interval between two supernovae).
For each cluster, the interval between two \emph{successive} supernova is monitored, 
within the numerical resolution $\mathrm{d}\log\Delta t=0.05$. Colour codes for different numbers of stars
$N_{\star}$, logarithmically sampled between 10 and 500 
(purple = 10, blue = 27, green = 71, orange = 189, red = 500).}
\label{fig:delta_shocks_2}
\end{figure}

Knowledge of the time distribution of supernovae is important to acceleration
in superbubbles, because, depending on the typical interval between shocks, 
accelerated particles may or may not remain within the bubble between two 
supernovae explosions, and thus experience repeated acceleration\footnote{
Note that this will also strongly depend on the initial energy of
the particles: the higher the energy they have gained from one shock,
the sooner they will escape the bubble, and hence the smaller chance they
have to be reaccelerated by a subsequent shock.}. 
We thus monitor the time interval $\Delta t_{SN}$ between 
two \emph{successive} supernovae. 
The result is shown in Fig.~\ref{fig:delta_shocks_2}.
We note that (1)~the most probable time interval between
two shocks is simply the average time between two supernovae 
$\bar{\Delta}t_{\mathrm{SN}}={\Delta t_{\mathrm{OB}}^{\star}} / {N_{\star}}\:;$
and (2)~when time intervals are normalised by this quantity,
all distributions have the same shape independently of the number
of stars (apart from very low numbers of stars).

\begin{figure}[!t]
\resizebox{\hsize}{!}{\includegraphics{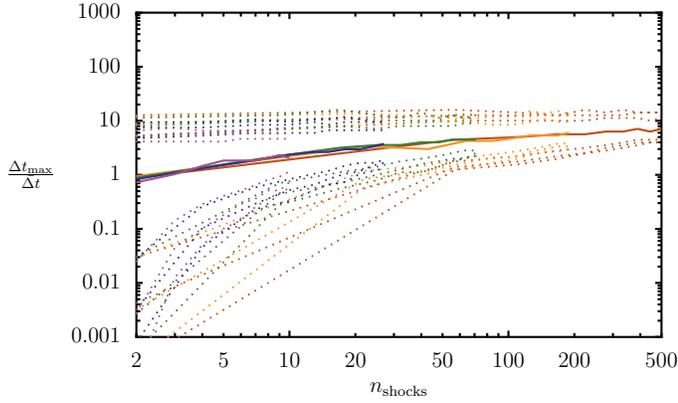}}
\caption{Maximum time interval between two successive shocks in a sequence of $n$ successive shocks 
(normalised to the average interval between two supernovae).
Solid curves correspond to the most frequent value of $\Delta t_{\mathrm{max}}$
(i.e., maxima of the curves in Fig.~\ref{fig:delta_shocks_2}). 
Dotted lines indicate the envelope of the distribution (they correspond to a decrease 
in the maximum value by a factor of 10, 100, 1000). 
Colours code the number of stars $N_{\star}$ in the same way as in Fig.~\ref{fig:delta_shocks_2}
(note that $N_{\star}$ coincides with the maximum number of successive
shocks $n$ for which data are available). }
\label{fig:delta_shocks_MAX}
\end{figure}

To investigate the probability of acceleration by 
\emph{many} successive shocks, we now compute the 
maximum time $\Delta t_{\mathrm{max}}$ that a particle has to wait 
\emph{within} a sequence of $n$ \emph{successive} shocks. 
Only particles whose escape time is longer than this value may experience 
acceleration by $n$ shocks. As previously, all distributions have the same shape 
once time intervals are normalised by $\bar{\Delta}t_{\mathrm{SN}}$, and are very peaked,
but now the most probable value of $\Delta t_{\mathrm{max}}$ 
is a few times longer than the average value 
(the more successive shocks we consider, the higher the probability
of obtaining an unusually long time interval between any two of them). 
This is summarised in Fig.~\ref{fig:delta_shocks_MAX}, which shows
the most probable value of $\Delta t_{\mathrm{max}}$ as a function of the number of successive shocks. 
We note that $\Delta t_{\mathrm{max}}$ may reach 10~times $\bar{\Delta}t_{\mathrm{SN}}$, 
and that it is an imprecise indicator when $N_{\star}$ and $n$ are low.

\subsection{Average cosmic-ray spectra}
\label{sec:res_spectra}

\subsubsection{General trends}
\label{sec:trends}

\begin{figure*}[!t]
\centering
~\\
\includegraphics[width=9cm]{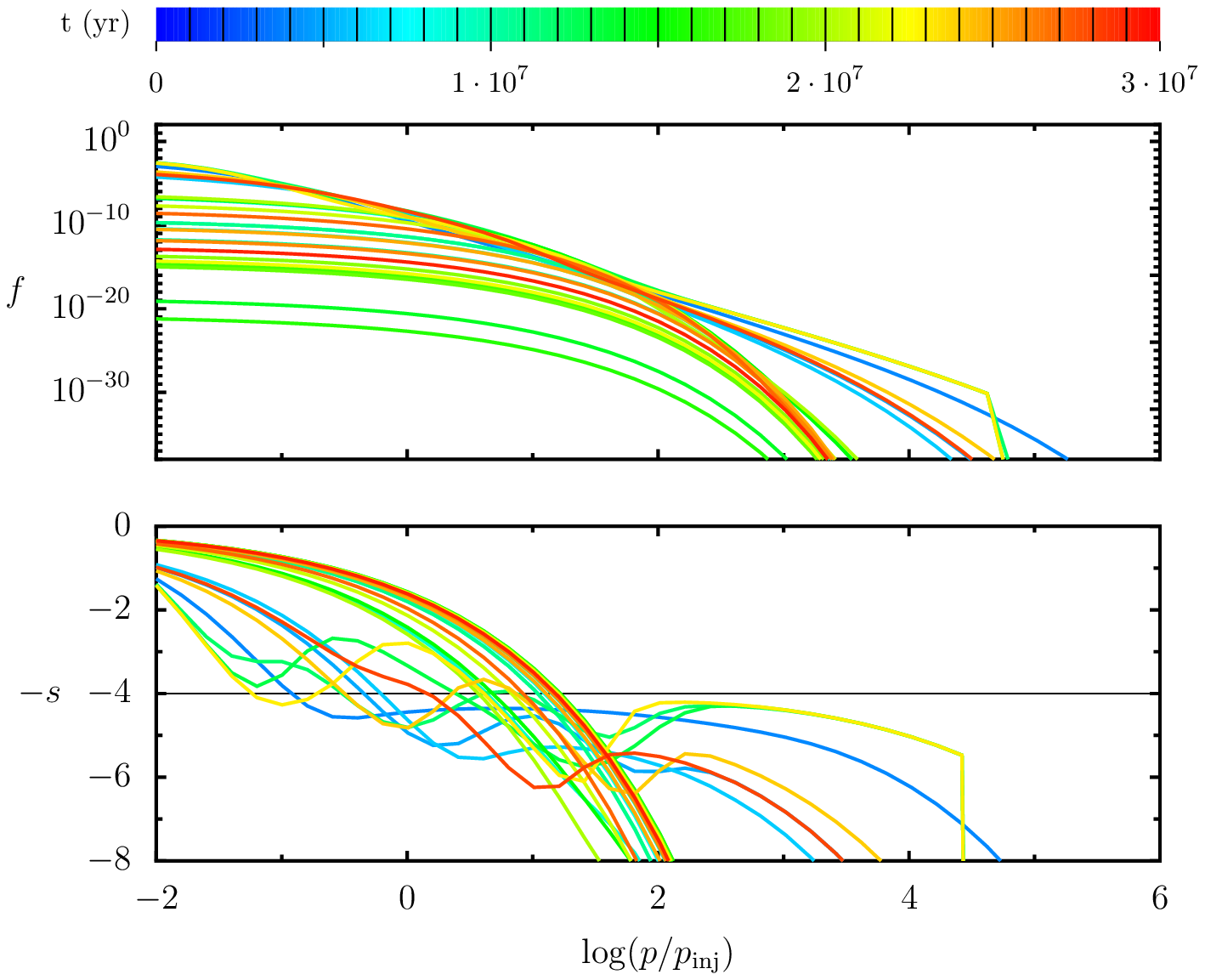}
\includegraphics[width=9cm]{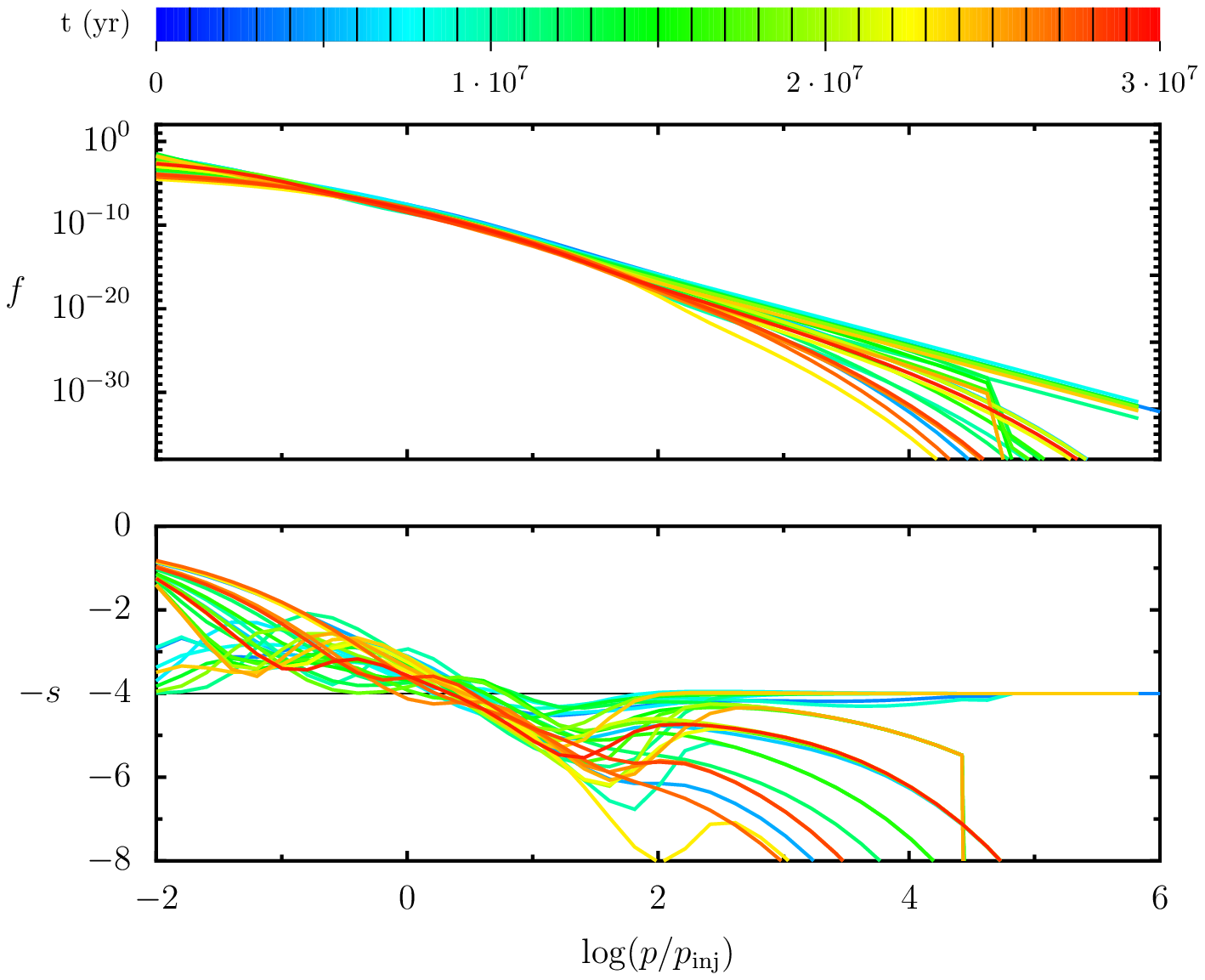}
\includegraphics[width=9cm]{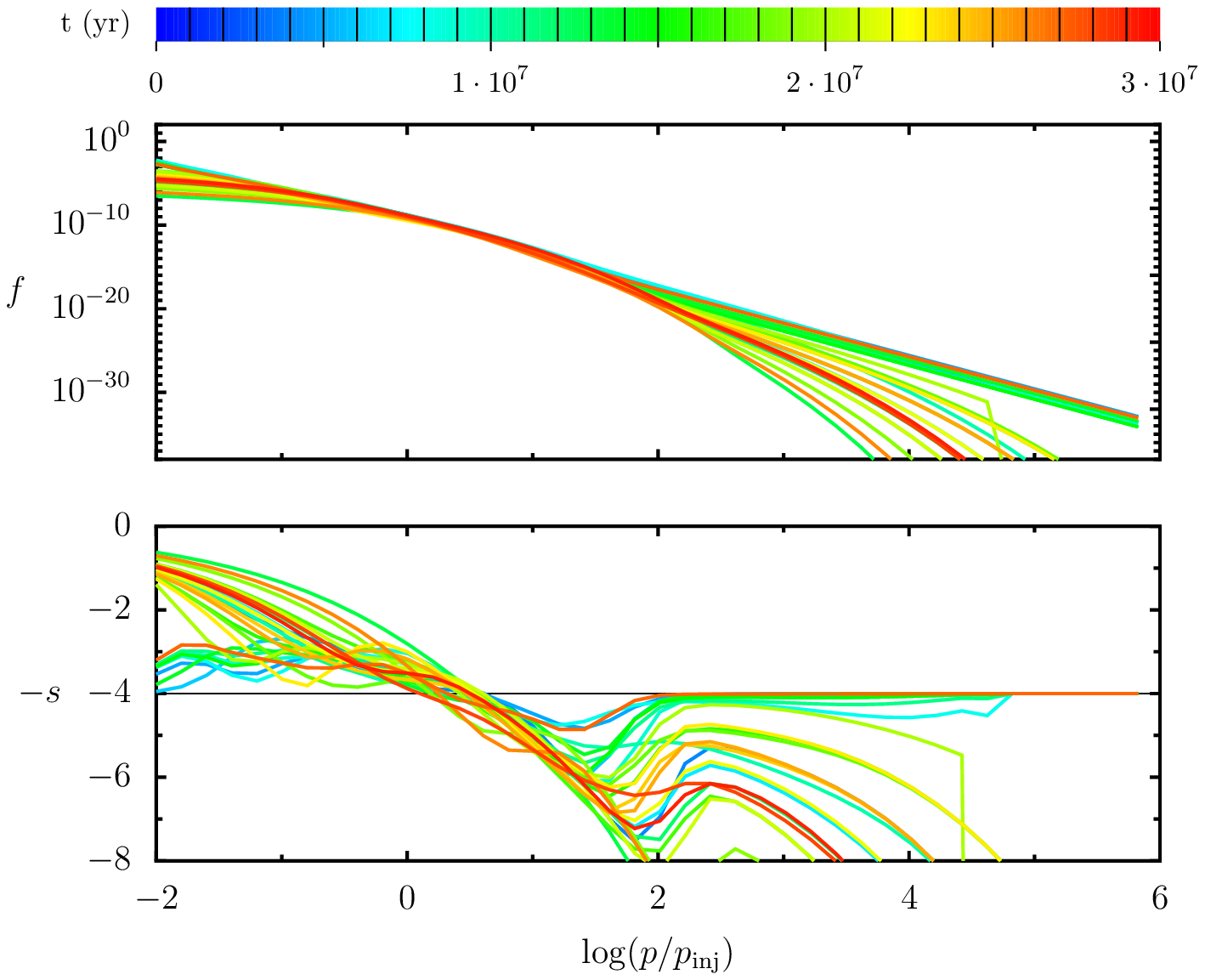}
\includegraphics[width=9cm]{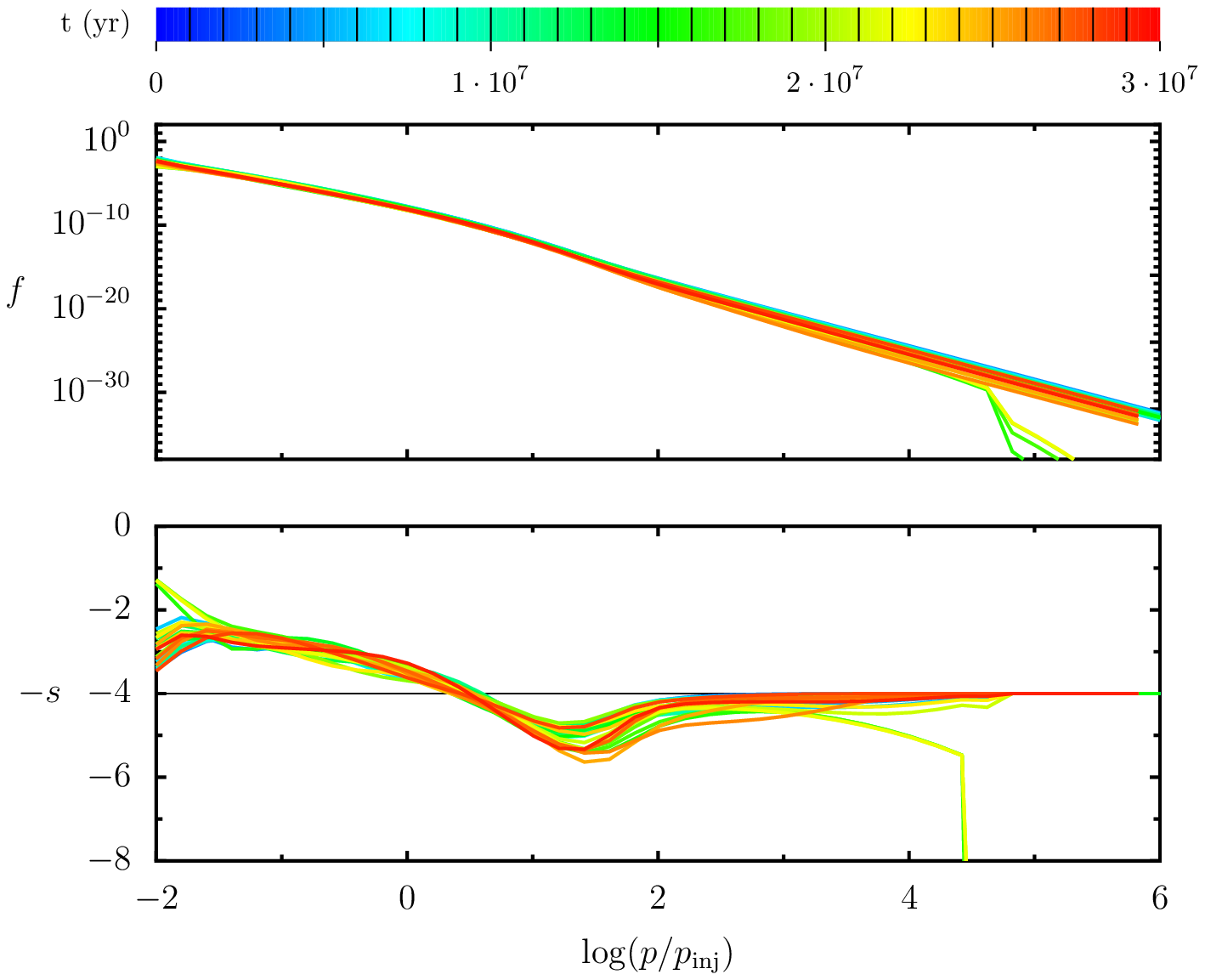}
\includegraphics[width=9cm]{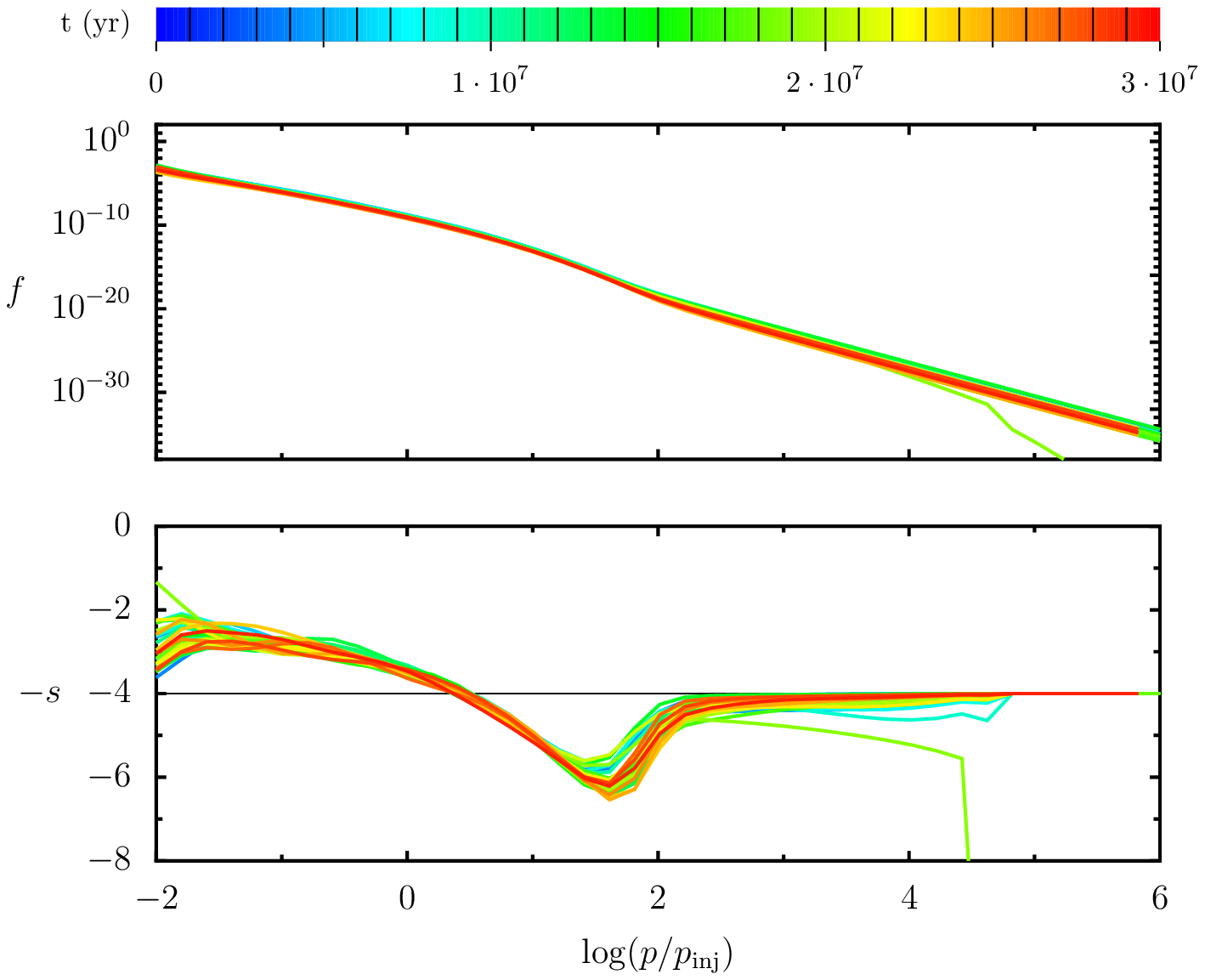}
\includegraphics[width=9cm]{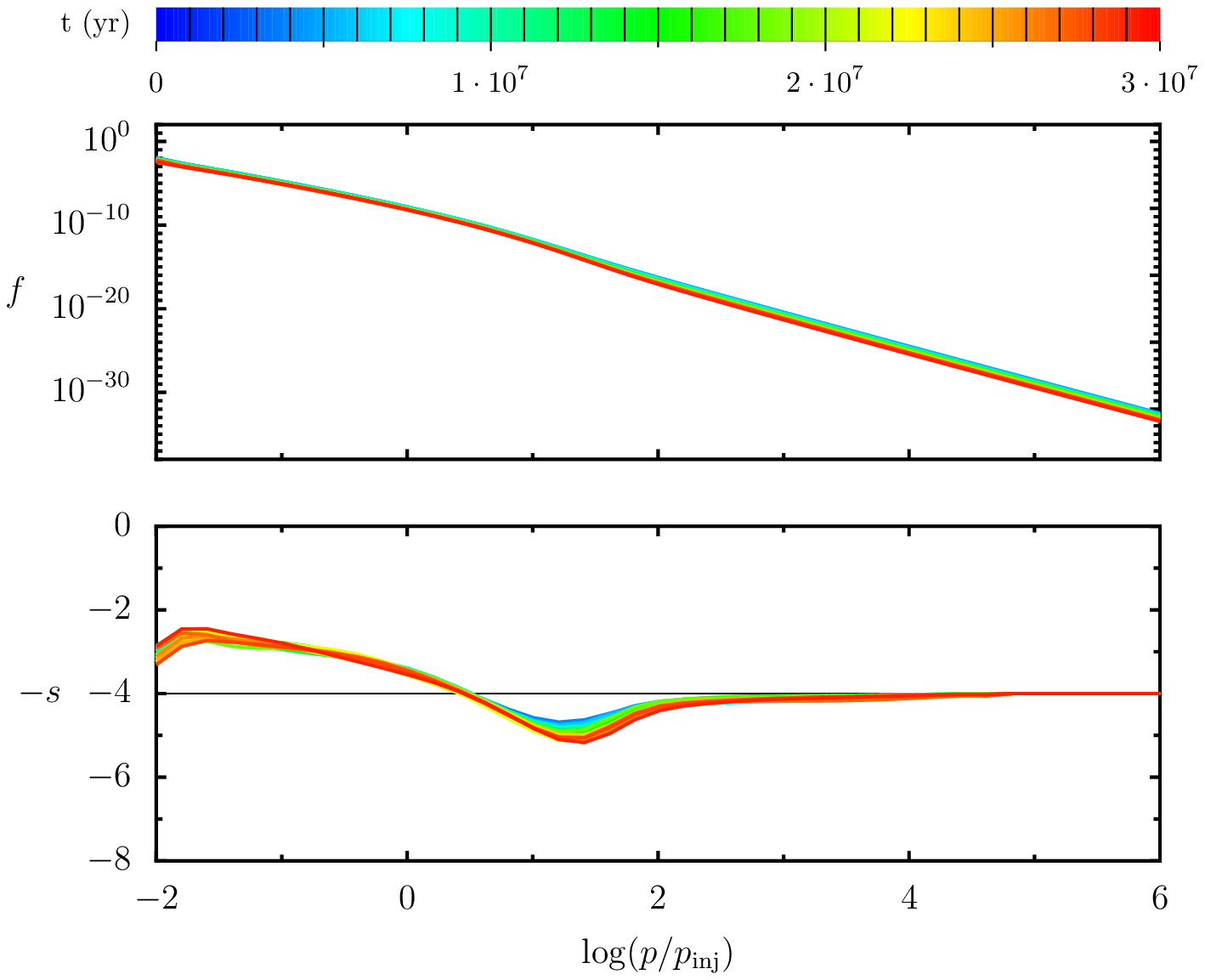}
\caption{Sample results of average spectra of cosmic rays inside the superbubble.
The particles spectrum $f$ and its logarithmic slope $s=\mathrm{d}\log f/\mathrm{d}\log p$
are plotted versus momentum $p$. The size of the cluster is $N_{\star}=10$ (left) 
and $N_{\star}=100$ (right). The number of samplings rises from top to bottom: $N=10,100,1000$. 
Other parameters are $q=5/3$, $B=10\:\mathrm{\mu G}$, $\eta_{T}=1$, $\lambda_{\mathrm{max}}=10\:\mathrm{pc}$, $x_{\mathrm{acc}}=40\:\mathrm{pc}$, $n=10^{-2}\:\mathrm{cm^{-3}}$.}
~\\
~\\
\label{fig:sample_results_rem}
\end{figure*}

\begin{figure*}[!t]
\centering
~\\
\includegraphics[width=9cm]{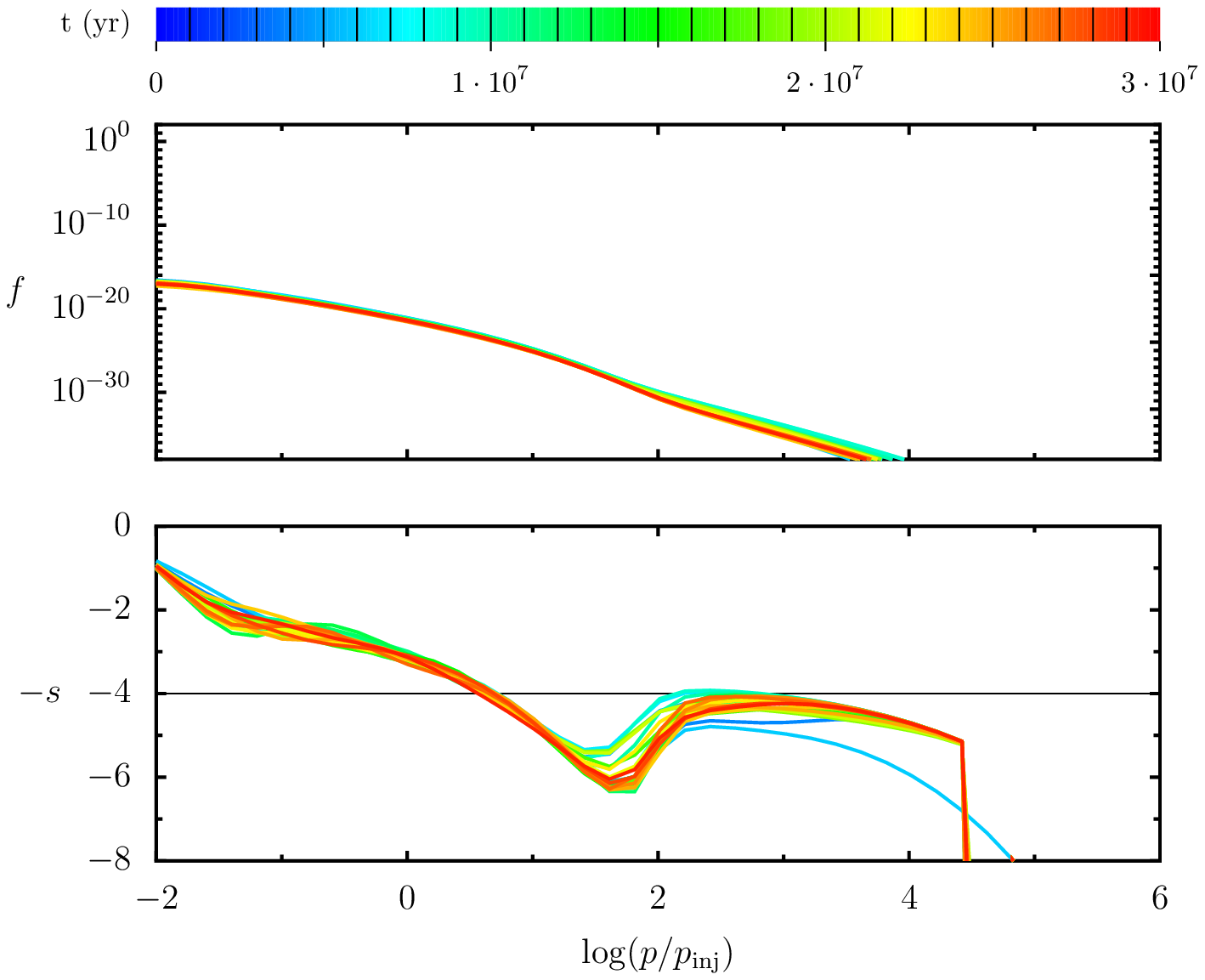}
\includegraphics[width=9cm]{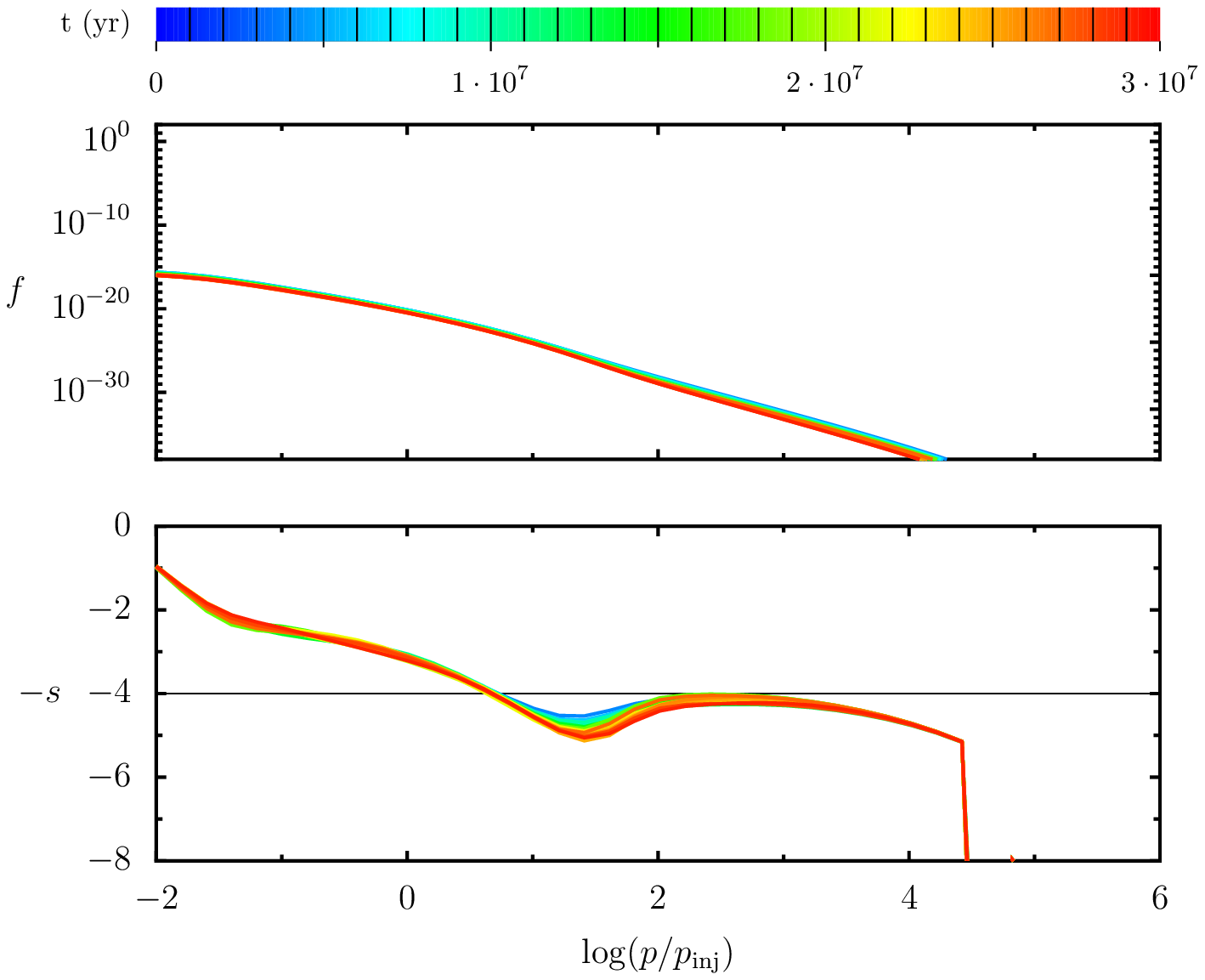}
\caption{Sample results of average spectra of cosmic rays escaping the superbubble.
The particles spectrum $f$ per unit time and its logarithmic slope $s=\mathrm{d}\log f/\mathrm{d}\log p$
are plotted versus momentum $p$. The size of the cluster is $N_{\star}=10$
(left) and $N_{\star}=100$ (right). The number of samplings is $N=1000$.
Other parameters are as in Fig.~\ref{fig:sample_results_rem}: $q=5/3$, $B=10\:\mathrm{\mu G}$, $\eta_{T}=1$, $\lambda_{\mathrm{max}}=10\:\mathrm{pc}$, $x_{\mathrm{acc}}=40\:\mathrm{pc}$, $n=10^{-2}\:\mathrm{cm^{-3}}$.}
~\\
~\\
\label{fig:sample_results_esc}
\end{figure*}

Proton spectra for clusters of two different sizes inside a typical superbubble 
are shown in Fig.~\ref{fig:sample_results_rem}.
For a given sample, we observe a strong intermittency during the
cluster lifetime (from blue to red), especially at early times. Nevertheless,
we clearly see convergence to an average spectrum as we increase the
number~$N$ of samples (from top to bottom). Comparing results for 10 and
100 stars (left and right), we see that what actually matters
is the total number of supernovae $N\times N_{\star}$. 
The limit spectrum exhibits a distinctive two-part shape, with a transition from a hard regime 
(flat spectrum, of slope $s<4$) to a soft regime (steep spectrum, of slope $s\geq4$). 
We also show the escaping spectra in Fig.~\ref{fig:sample_results_esc}. 
We see that they have the same overall shape, but are a bit harder 
(as highly energetic particles escape first) and of much lower normalization.

Hard spectra at low energies are produced by the combined effects of acceleration 
by supernova shocks (Fermi~1) and reacceleration by turbulence (Fermi~2). 
Soft spectra at high energies are mostly shaped by escape, which preferentially 
removes highly energetic particles. The transition energy is controlled 
by a balance between reacceleration and escape timescales, and thus
depends on the superbubble parameters.

\subsubsection{Parametric study}
\label{sec:parameters}

\begin{figure*}[!t]
\sidecaption
\includegraphics[height=6.5cm]{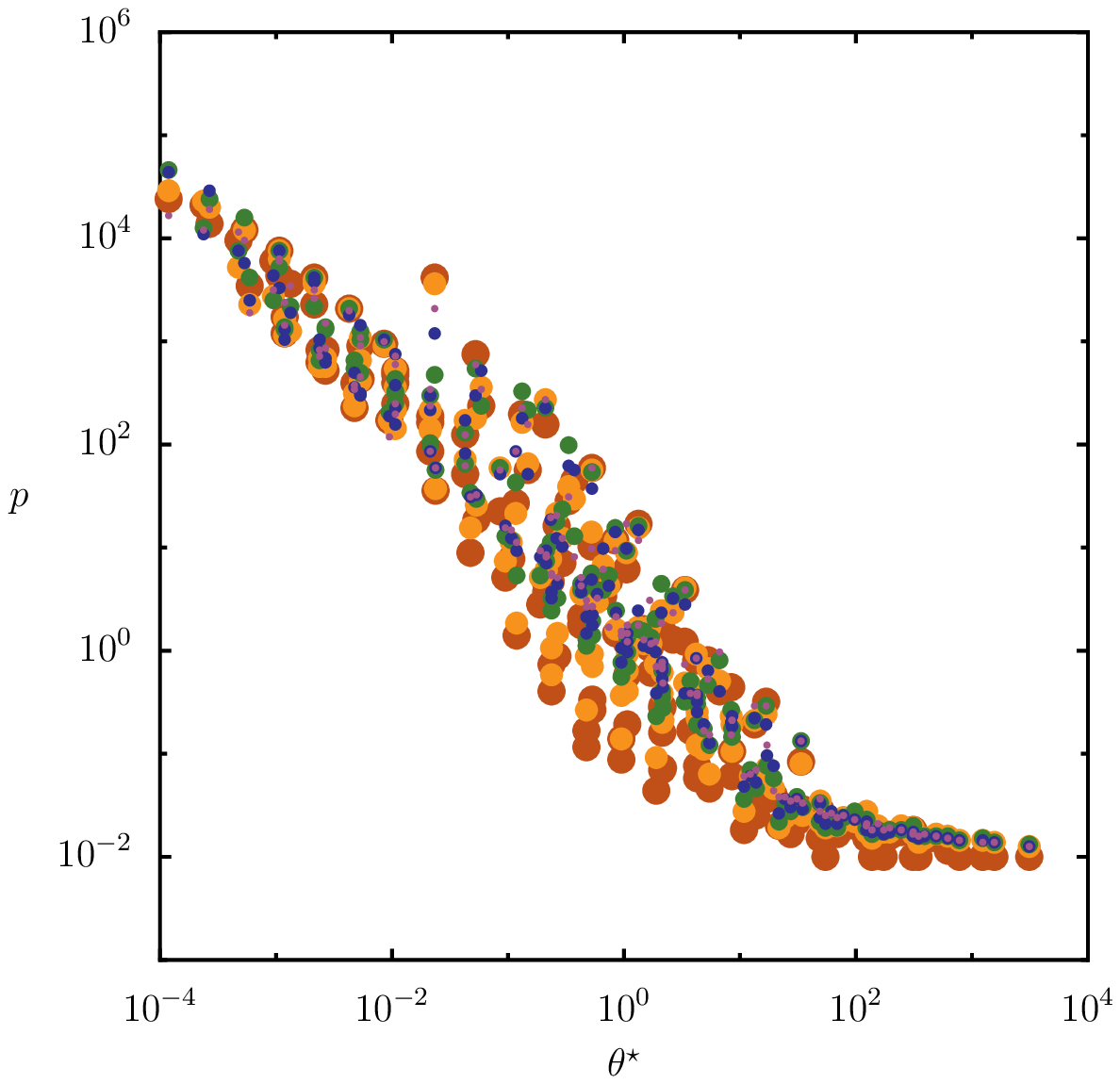}
\includegraphics[height=6.5cm]{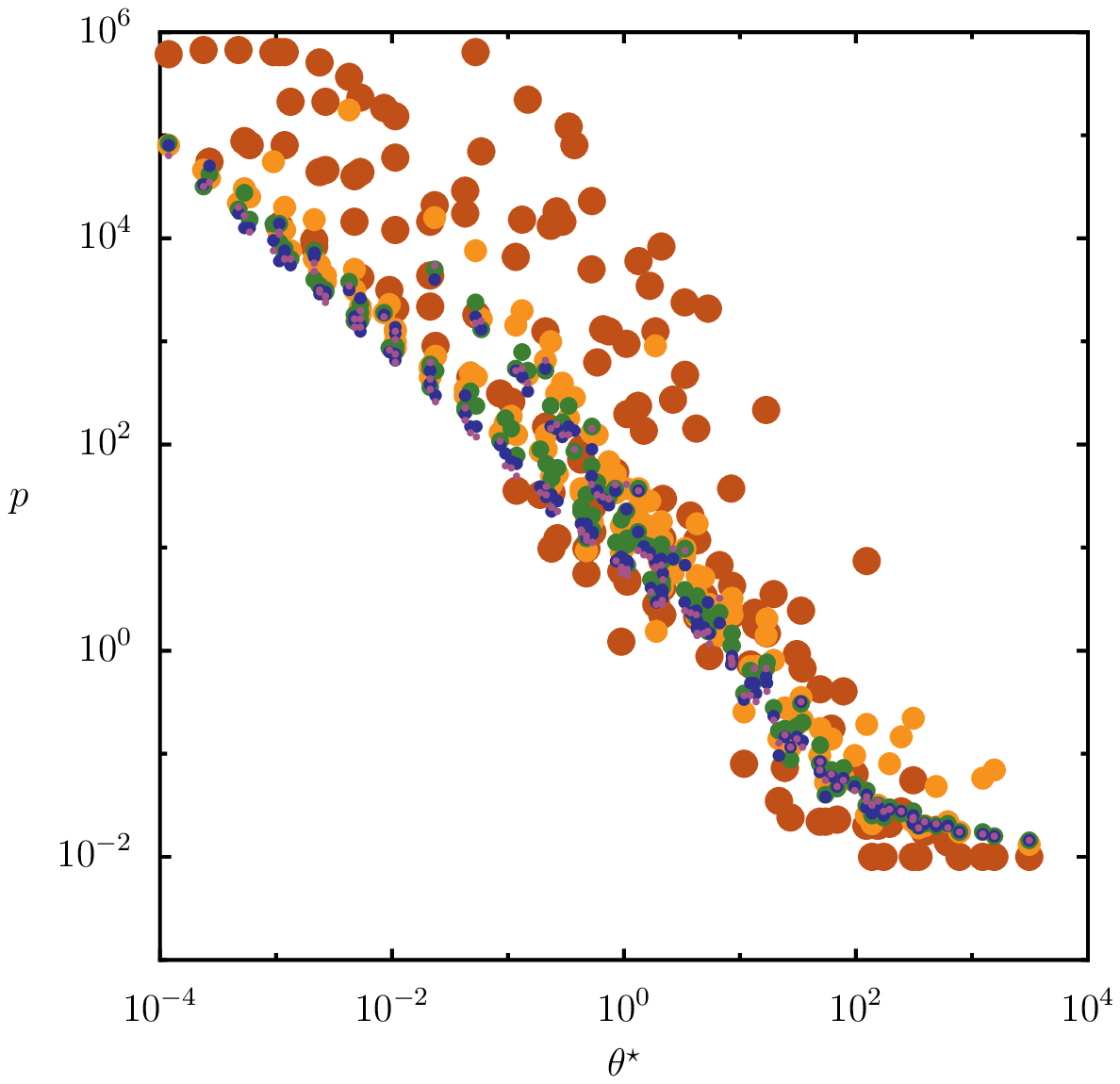}
\caption{Hard-soft transition momentum as a function of $\theta^{\star}$ and
$N_{\star}$.
The transition momentum is defined as the momentum up to which the
particle spectrum may have index lower than a given threshold: 
$s=3$ at the left, $s=4$ at the right. 
$\theta^{\star}$ is a dimensionless parameter defined by Eq.~(\ref{eq:Theta_def}). 
The number of stars $N_{\star}=$10,30,70,200,500 
is coded by both dot sizes and dot colours. 
Momentum resolution is 10~bins per decompression shift, 
that is $\simeq50$~bins per decade.\newline\newline\newline}
~\\
\label{fig:transition_momentum}
\end{figure*}

\begin{figure*}[!t]
\sidecaption
\includegraphics[height=6.5cm]{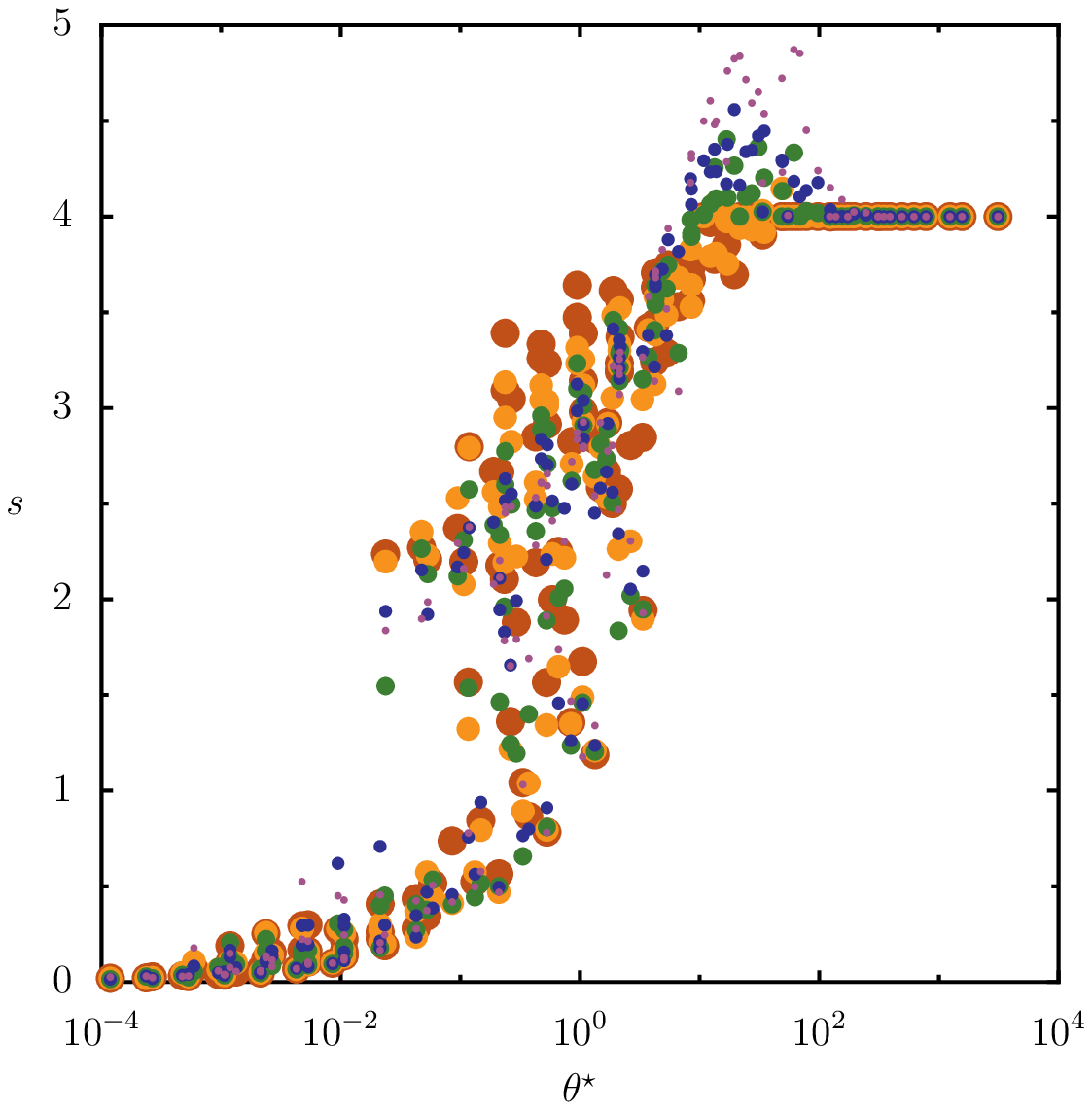}
\includegraphics[height=6.5cm]{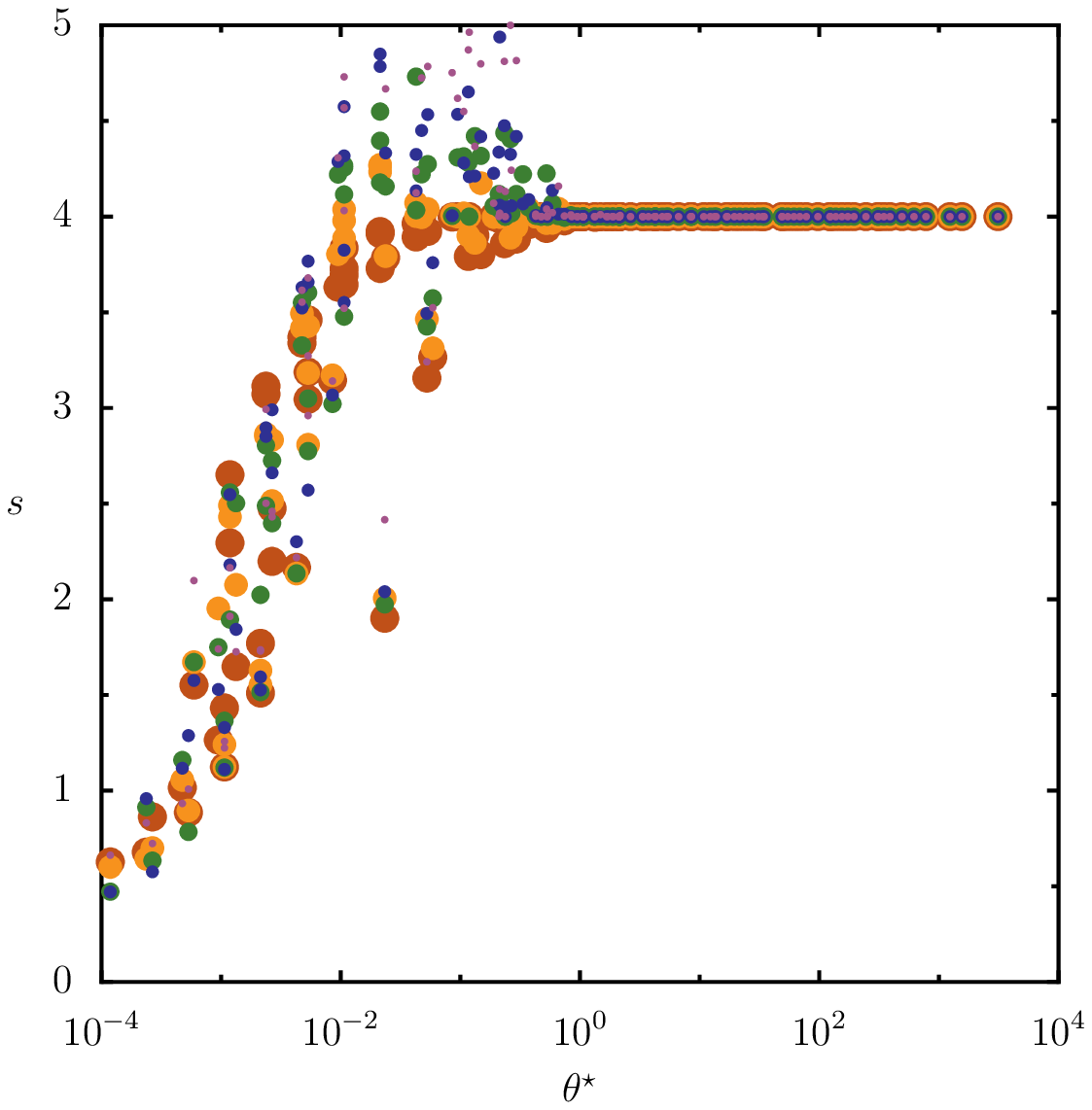}
\caption{Lowest slope as a function of $\theta^{\star}$ and $N_{\star}$.
The plots show the lowest slope (corresponding to the hardest spectrum) 
reached at a given momentum: 
$p=1\:\mathrm{GeV}$ at the left, $p=1\:\mathrm{TeV}$ at the right. 
$\theta^{\star}$ is a dimensionless parameter defined by Eq.~(\ref{eq:Theta_def}). 
The number of stars $N_{\star}=$10,30,70,200,500
is coded by both dot sizes and dot colours. 
Momentum resolution is 10~bins per decompression shift, 
that is $\simeq50$~bins per decade.\newline\newline\newline\newline}
~\\
\label{fig:smallest_slope}
\end{figure*}

For each cluster, we must define eight parameters $N_{\star}$, $r$, $q$, $\eta_{T}$, $B$, 
$\lambda_{\mathrm{max}}$, $n$, and $x_{\mathrm{acc}}$, which are more or less constrained. 
We sample the size of the cluster roughly logarithmically between 10 stars and 500 stars , 
i.e. $N_{\star}=$10, 30, 70, 200, 500. 
We consider only strong supernova shocks of $r=4$.
We compare the classical turbulence indices $q=5/3$
(Kolmogorov cascade, K41) and $q=3/2$ (Kraichnan cascade, IK65). 
We consider two different scenarios for the magnetic field: if a turbulent dynamo
is operating then $B\simeq10\:\mathrm{\mu G}$ and $\delta B\gg B$,
so that $\eta_{T}\simeq1$ \citep{Bykov2001b}; if not, then
because of the bubble expansion $B\simeq1\:\mathrm{\mu G}$ and $\delta B<B$
(if $\delta B=B/2$, then $\eta_{T}=0.2$). 
The external scale of the turbulence $\lambda_{\mathrm{max}}$ is at least of the order 
of the distance $d_{\star}$ between two stars in the cluster, which, for
a typical OB~association radius of 35~pc (e.g.,~\citealt{Garmany1994a}),
and assuming uniform distribution (a quite crude approximation), is
\begin{equation}
\label{eq:d_stars}
d_{\star} \simeq \frac{56\:\mathrm{pc}}{N_{\star}^{1/3}}\:,
\end{equation}
which is 26, 12 and 7~pc for 10, 100 and 500~stars respectively.
However, $\lambda_{\mathrm{max}}$ will be higher if turbulence is driven by supernova remnants, 
the radius of which increases roughly as 
\begin{equation}
\label{eq:r_SNR}
r_{\mathrm{SNR}} \simeq 38\:\mathrm{pc\:\left(\frac{t}{10^{4}\: yr}\right)^{{2}/{5}}}
\end{equation}
in the Sedov-Taylor phase inside a superbubble \citep{Parizot2004a}. 
Hence, we consider $\lambda_{\mathrm{max}}=10,20,40,80\:\mathrm{pc}$.
We consider the size of the acceleration region to be of the order of
the radius of a supernova remnant after our time-step $\mathrm{d}t=10\:000\:\mathrm{yr}$,
which is $x_{\mathrm{acc}}=40\:\mathrm{pc}$ according to Eq.~(\ref{eq:r_SNR}). 
However, in evolved superbubbles it might be higher, up to more than 100~pc, 
so we also try 80~pc and 120~pc. 
The density inside a superbubble is always low, and to assess its influence we perform simulations with 
$n=10^{-3}\:\mathrm{cm^{-3}}$, $n=5\times10^{-3}\:\mathrm{cm^{-3}}$, and $n=10^{-2}\:\mathrm{cm^{-3}}$.
This provides 720 different cases to run. 
And in each case, we have to set the number $N$ of samplings per cluster: 
convergence of average spectra typically requires $N_{\star}\times N\simeq10^4$, 
but the general trend is already clear as soon as $N_{\star}\times N\simeq10^3$,
so we simply take $N=10^3/N_{\star}$.

We thus had to perform many simulations to explore the parameter space.
However, interestingly, the effects of the 6~parameters 
relevant to stochastic acceleration and escape 
$q$, $\eta_{T}$, $B$, $\lambda_{\mathrm{max}}$, $n$, and $x_{\mathrm{acc}}$
can be summarized by a single parameter, the adimensional number
$\theta^{\star}$ introduced by \citet{Becker2006a}
\begin{equation}
\theta^{\star} = \frac{1}{{D_{p}^{\star}\: t_{\mathrm{esc}}^{\star}}}\:,
\label{eq:Theta_def}
\end{equation}
which, according to Eqs.~(\ref{eq:Dp*}) and~(\ref{eq:Tesc*}) varies as
\begin{equation}
\theta^{\star}\propto\eta_{T}^{-2}\: B^{2q-6}\:\lambda_{\mathrm{max}}^{2q-2}\: 
x_{\mathrm{acc}}^{-2}\: n\:.
\label{eq:Theta*}
\end{equation}
For standard turbulence indices, we have
\begin{equation}
\label{eq:Theta_ref}
\theta^{\star}\simeq
\left\{
\begin{array}{ll}
\frac{2}{\eta_{T}^2}\left(\frac{B}{10\:\mathrm{\mu G}}\right)^{-\frac{8}{3}}\left(\frac{\lambda_{\mathrm{max}}}{10\:\mathrm{pc}}\right)^{\frac{4}{3}}\left(\frac{x_{\mathrm{acc}}}{40\:\mathrm{pc}}\right)^{-2}\left(\frac{n}{10^{-2}\:\mathrm{cm^{-3}}}\right) & q=5/3\\
\frac{10^{-2}}{\eta_{T}^2}\left(\frac{B}{10\:\mathrm{\mu G}}\right)^{-3}\left(\frac{\lambda_{\mathrm{max}}}{10\:\mathrm{pc}}\right)\,\left(\frac{x_{\mathrm{acc}}}{40\:\mathrm{pc}}\right)^{-2}\left(\frac{n}{10^{-2}\:\mathrm{cm^{-3}}}\right) & q=3/2
\end{array}
\right.\:.
\end{equation}
For all the possible superbubble parameters considered here,
$\theta^{\star}$ ranges from $10^{-4}$ to $10^{+4}$. Since we consider
only strong supernova shocks of $r=4$, the single remaining parameter is
the number of stars $N_{\star}$ (represented by dots of different
colours and sizes in subsequent plots), which has a weaker impact on our results. 

To characterize the spectra of accelerated particles, we use two indicators, 
which are plotted in Figs.~\ref{fig:transition_momentum} and~\ref{fig:smallest_slope}. 
We checked that the results are independent of the resolution, 
provided that there is at least a few bins per decompression shift.
The residual variability seen originates mostly in the simulation procedure itself, 
which is based on random samplings.
In Fig.~\ref{fig:transition_momentum}, we show the momentum of
transition from hard to soft regimes, defined as the maximum momentum
up to which the slope may be smaller than a given value (3 or 4 here).
Above this momentum, the slope always remains greater than this value. 
Below this momentum, the slope can be as low as 0,
meaning that particles pile-up from injection -- but we note that it can also
happen to be $\geq4$ at a particular time in a particular cluster sample, 
since distributions are highly variable. 
As $\theta^{\star}$ increases, the transition momentum falls exponentially
from almost the maximum momentum considered (a fraction of PeV)
to the injection momentum (10~MeV). For rule-of-thumb calculations,
one can say that the slope can be $<3$ up to $p=1/\theta^{\star}$~GeV.
In Fig.~\ref{fig:smallest_slope}, we show the shallowest slope (corresponding
to the hardest spectrum) obtained at a fixed momentum (1~GeV and
1~TeV here). As $\theta^{\star}$ increases, the lowest slope rises from 0 
(which is possible in the case of stochastic reacceleration) to 4 
(the canonical value for single regular acceleration in the test particle case).
As expected, the critical $\theta^{\star}$ between hard and soft regimes 
decreases as we increase the reference momentum: the break occurs 
around $\theta^{\star}=10$ for $p=1\:\mathrm{GeV}$,
and around $\theta^{\star}=0.01$ for $p=1\:\mathrm{TeV}$.

This overall behaviour can be explained by noting that $\theta^{\star}$ is roughly
the ratio of the reacceleration time to the escape time. 
Low~$\theta^{\star}$ are obtained when reacceleration is faster than escape, 
allowing Fermi processes to produce hard spectra up to high energies, 
as particles become reaccelerated by shocks and/or turbulence.
In contrast, high~$\theta^{\star}$ are obtained when escape is faster than reacceleration, 
resulting in quite soft in-situ spectra,
as particles escape immediately after being accelerated by a supernova shock.
The case $\theta^{\star}=1$ corresponds to a balance between gains and losses,
in the particular case of which the spectral break occurs around 10~GeV for $s>4$, 
and around 1~GeV for $s>3$. 


\section{Application}
\label{sec:application}


\subsection{A selection of massive star regions}
\label{sec:selection}

We gathered the physical parameters of some well observed massive star clusters and their associated superbubbles. The reliability and the completeness of the data were our main selection criteria. The parameters useful for our study are: the cluster composition (number of massive stars), age, distance, size, and the superbubble size and density. We note that we are biased towards young objects, since older ones are more difficult to isolate because of their large extensions and sequential formations. Information about density is sometimes unavailable. The density can span several orders of magnitude, usually between $10^{-2}$ and $10 \ \rm{cm^{-3}}$ in the central cluster \citep{Torres2004a}, and between $10^{-3}$ and $10^{-1} \ \rm{cm^{-3}}$ in the superbubble \citep{Parizot2004a}. If X-ray observations are available, it can be indirectly estimated from the thermal X-ray spectrum, given the plasma temperature and the column density along the line of sight. In the case of a complete lack of data, we accept a mean density of between $5\times10^{-3}\ \rm{cm^{-3}}$ and $5\times10^{-2}\ \rm{cm^{-3}}$. Unfortunately, the magnetic field parameters can not be directly measured, so that we consider different limiting scenarios: $B=1\:\mu\rm{G}$ and $\eta_T=0.2$ if the turbulence is low, and $B=10\:\mu\rm{G}$ and $\eta_T=1$ if the turbulence is high. In each case, we compare our results for turbulence indices $q=5/3$ and $q=3/2$. The maximal scale of the turbulence $\lambda_{\rm max}$ may be taken to be as small as the size of the stellar cluster (especially in the case where few supernovae have already occurred), or as large as the superbubble itself.

These quantities are used to estimate the key parameter $\theta_{\star}$ in each of the selected objects using Eq.~(\ref{eq:Theta_ref}). All the parameters and results are summarised in Table~1. Before discussing the implications of these values, we provide details of the selected regions in the following two sections, regarding clusters found in our Galaxy and in the Large Magellanic Cloud (LMC), respectively.

\subsubsection{Galaxy}
\label{sec:CGal}

We selected 6~objects in the Galaxy.

\begin{itemize} 

\item Cygnus region: in this region we identify two distinct objects, the clusters Cygnus OB1 and OB3, which have blown a common superbubble, and the cluster Cygnus OB2. We note that the latter was detected at TeV energies by Hegra  \citep{Aharonian2005b} as an extended source (TeV J2032+4130), and by Milagro \citep{Abdo2007a}, as extended diffuse emission and at least one source (MGRO J2019+37). A supershell was also detected around the Cygnus X-ray superbubble, which may have been produced by a sequence of starbursts, Cygnus OB2 being the very last.

\item Orion OB1: this association consists of several subgroups \citep{Brown1999a}, the age of 12~Myrs selected here corresponds to the oldest one (OB1a).

\item Carina nebula: this region is one of the most massive star-forming regions in our Galaxy. It contains two massive stellar clusters, Trumpler~14 and Trumpler~16 \citep{Smith2000a}, of cumulative size of approximately 10~pc.

\item Westerlund~1:  this cluster is very compact although it harbours hundreds of massive stars. The size of the superbubble is uncertain, and we assume here the value of 40~pc reported by \cite{Kothes2007a} for the HI shell surrounding the cluster. We note that Westerlund~1 was detected by HESS \citep{Ohm2009a}.

\item Westerlund~2: the distance to this cluster remains a matter of debate (see the discussion in \citealt{Aharonian2007a}), and we adopt here the estimate of \cite{Rauw2007a}, using it to re-evaluate the size obtained by \cite{Conti2004a}. We assume that the giant HII region RCW49 of size 100~pc is the structure blown by Westerlund~2. \cite{Tsujimoto2007a} provided a spectral fit of the diffuse X-ray emission from RCW49, from which we deduce a density $\sim 1.5\times10^{-3} \ \rm{cm^{-3}}$. We note that Westerlund~2 was detected by HESS \citep{Aharonian2007a}. 

\end{itemize}

\subsubsection{Large Magellanic Cloud}
\label{sec:CLMC}

We selected 3~objects in the LMC. All density estimates here have been derived from observations of diffuse X-ray emission. At the distance of the LMC, these observations usually cover the entire structure, so that the density deduced is an average over the OB association and the ionised region around it.

\begin{itemize} 
\item DEML~192: this region harbours two massive star clusters, LH~51 and~54 \citep{Lucke1970a}. We deduced the spatial extensions of both clusters from \cite{Oey1998a}, but these are probably overestimates, because the edges of the clusters are not clearly defined. 

\item 30~Doradus: this region is quite complex as can be seen from Chandra observations \citep{Townsley2006a}. In particular, the superbubble extension is difficult to estimate precisely. We decided to assume the value given for the 30~Doradus nebula by \cite{Walborn1991a}. The extension of the star cluster may be larger than the core which harbours several thousands of stars \citep{Massey1998a}. The core size is $\le 10$~pc \citep{Massey1998a}, it is even estimated to be $\sim 2$~pc by \cite{Walborn1991a}. The number of massive stars in R136 depends on the cluster total mass, estimated to be between $5\times10^4 \ \rm{M_{\odot}}$ and $2.5\times10^5 \ \rm{M_{\odot}}$. Using a Salpeter IMF, one finds that $N_{\star} (M > 8 M_{\odot}) \simeq 400-2700$. We note that the stellar formation in 30~Doradus was sequential and started more than 10~Myrs ago \citep{Massey1998a}. 

\item N11: this giant HII region harbours several star clusters LH9, LH10, LH13, and LH14, probably produced as a sequence of starbursts \citep{Walborn1999a}. Here we mostly consider the star cluster LH9 at the center of N11 and the shell encompassing it (shell~1 in \citealt{Mac-Low1998a}). LH10 is a younger star cluster with an estimated age of 1~Myr \citep{Walborn1999a} in which no supernova has yet occurred. The other clusters are less powerful. 

\end{itemize}

\begin{table*}[t]
\label{tab:obs}
\caption{Physical parameters for well observed massive-star forming regions in the Galaxy and in the LMC.}
\begin{center}
\begin{tabular}{|m{1.9cm}|m{1.2cm}m{0.8cm}m{1.3cm}m{1.cm}|m{1.5cm}m{1.5cm}|m{1.cm}m{1.cm}m{1.cm}m{1.cm}|}
\hline & \multicolumn{4}{c|}{Cluster} & \multicolumn{2}{c|}{Superbubble} & \multicolumn{4}{c|}{$\theta_{\star}^\mathrm{(d)}$} \\ \hline 
Name & $N_{\star}^\mathrm{(a)}$ & Age (Myr) & Distance (kpc) & Size$^\mathrm{(b)}$ (pc) & Size$^\mathrm{(b)}$ $\quad\quad$(pc) & Density$^\mathrm{(c)}$ ($\rm cm^{-3}$) & B=$1\mu\rm{G}$ $q=5/3$ & B=$1\mu\rm{G}$ $q=3/2$ & B=$10\mu\rm{G}$ $q=5/3$ & B=$10\mu\rm{G}$ $q=3/2$ \\ \hline \hline
Cygnus OB1/3 & 38$^{(16)}$          & 2-6$^{(12)}$  & 1.8$^{(19)}$       & 24                     & 80-100$^{(14)}$   & $0.01?$                 
	& $5.10^4$\textcolor{white}{-}$5.10^5$  & $4.10^2$\textcolor{white}{-}$3.10^3$ & $4.10^{0}$\textcolor{white}{-}$4.10^{1}$  & $2.10^{-2}$\textcolor{white}{-}$1.10^{-1}$ \\ \hline  
Cygnus OB2    & 750$^{(5)}$          & 3-4$^{(12)}$  & 1.4-1.7$^{(10)}$ & 60$^{(11)}$      & 450?$^{(5)}$        & 0.02$^{(5)}$          
	& $2.10^4$\textcolor{white}{-}$2.10^5$  & $9.10^1$\textcolor{white}{-}$7.10^2$ & $1.10^0$\textcolor{white}{-}$2.10^1$  & $4.10^{-3}$\textcolor{white}{-}$3.10^{-2}$ \\ \hline 
Orion OB1       & 30-100$^{(3)}$     & 12$^{(2)}$     & 0.45$^{(2)}$       & 10$^{(2)}$         & 140x300$^{(3)}$  & 0.02-0.03$^{(4)}$  
	& $3.10^3$\textcolor{white}{-}$2.10^6$  & $4.10^1$\textcolor{white}{-}$7.10^3$ & $3.10^{-1}$\textcolor{white}{-}$2.10^2$  & $1.10^{-3}$\textcolor{white}{-}$3.10^{-1}$ \\ \hline  
Carina nebula  & ?                          & 3$^{(23)}$      & 2.3$^{(8)}$         & 20                     & 110$^{(23)}$        & $0.01?$                
	& $2.10^4$\textcolor{white}{-}$2.10^6$  & $1.10^2$\textcolor{white}{-}$7.10^3$ & $1.10^0$\textcolor{white}{-}$1.10^2$  & $5.10^{-3}$\textcolor{white}{-}$3.10^{-1}$ \\ \hline 
Westerlund 1   & 450$^{(1)}$          & 3.3$^{(1)}$    & 3.9$^{(13)}$       & 1$^{(1)}$           & 40?$^{(13)}$       & $0.01?$                
	& $2.10^3$\textcolor{white}{-}$3.10^6$  & $5.10^1$\textcolor{white}{-}$2.10^4$ & $2.10^{-1}$\textcolor{white}{-}$3.10^2$  & $2.10^{-3}$\textcolor{white}{-}$8.10^{-1}$ \\ \hline
Westerlund 2   &14$^{(21)}$           & 2$^{(21)}$     & 8$^{(21)}$          & 1$^{(6)}$           & 100$^{(21,6)}$     & 0.0015$^{(24)}$   
	& $1.10^2$\textcolor{white}{-}$5.10^4$  & $2.10^0$\textcolor{white}{-}$2.10^2$ & $9.10^{-3}$\textcolor{white}{-}$4.10^0$  & $1.10^{-4}$\textcolor{white}{-}$1.10^{-2}$ \\ \hline \hline 
DEM L192       & 135                        & 3$^{(20)}$    & 50                       & 60$^{(20)}$       & 120x135$^{(9)}$  & 0.03$^{(7)}$          
	& $3.10^5$\textcolor{white}{-}$1.10^6$  & $2.10^3$\textcolor{white}{-}$4.10^3$ & $2.10^1$\textcolor{white}{-}$9.10^1$  & $6.10^{-2}$\textcolor{white}{-}$2.10^{-1}$ \\ \hline 
30 Doradus     & $>$ 400$^{(22)}$  & 2$^{(17)}$    & 50                       & 40$^{(25)}$       & 200$^{(25)}$        & 0.09$^{(27)}$        
	& $2.10^5$\textcolor{white}{-}$2.10^6$  & $1.10^3$\textcolor{white}{-}$7.10^3$ & $2.10^1$\textcolor{white}{-}$2.10^2$  & $6.10^{-2}$\textcolor{white}{-}$3.10^{-1}$ \\ \hline 
N11                 & 130                        & 5$^{(26)}$    & 50                       & 15x30$^{(18)}$ & 100x150$^{(9)}$  & 0.08$^{(15)}$        
	& $9.10^4$\textcolor{white}{-}$4.10^6$  & $9.10^2$\textcolor{white}{-}$2.10^4$ & $8.10^0$\textcolor{white}{-}$4.10^2$  & $3.10^{-2}$\textcolor{white}{-}$8.10^{-1}$ \\ \hline
\end{tabular}
\end{center}
\begin{list}{}{}
\item[$^{\mathrm{a}}$] $N_{\star}$ is the number of stars with mass $\ge 8 M_{\odot}$ (a Salpeter IMF has been assumed, expected for N11 where an index of 2.4 has been used).
\item[$^{\mathrm{b}}$] Sizes are either the diameter if the region is spherical, or the large and small semi-axis if the region is ellipsoidal. 
\item[$^{\mathrm{c}}$] The density is the Hydrogen nuclei density. 
\item[$^{\mathrm{d}}$] Estimates of $\theta_{\star}$ are calculated from Eq.~(\ref{eq:Theta_ref}), as explained in Sect.~\ref{sec:selection}. The range of values of $\theta_{\star}$ given for each object and for each magnetic configuration reflects uncertainties in the actual values of bubble density, accelerator size and turbulence scale. 
\item[References]: 
(1)~\cite{Brandner2008a}, (2)~\cite{Brown1994a}, (3)~\cite{Brown1995a}, (4)~\cite{Burrows1993a}, (5)~\cite{Cash1980a}, 
(6)~\cite{Conti2004a}, (7)~\cite{Cooper2004a}, (8)~\cite{Davidson1997a}, (9)~\cite{Dunne2001a}, (10)~\cite{Hanson2003a}, 
(11)~\cite{Knodlseder2000a}, (12)~\cite{Knodlseder2002a}, (13)~\cite{Kothes2007a}, (14)~\cite{Lozinskaya1998a}, (15)~\cite{Maddox2009a}, 
(16)~\cite{Massey1995a}, (17)~\cite{Massey1998a}, (18)~\cite{Naze2004a}, (19)~\cite{Nichols-Bohlin1993a}, (20)~\cite{Oey1998a}, 
(21)~\cite{Rauw2007a}, (22)~\cite{Selman1999a}, (23)~\cite{Smith2000a}, (24)~\cite{Tsujimoto2007a}, (25)~\cite{Walborn1991a}, 
(26)~\cite{Walborn1992a}, (27)~\cite{Wang1991a}. 
\end{list}
\end{table*}

\subsection{Discussion}

In Table~1, we can see that in all cases except for $q=3/2$, $B=10\:\mu \rm{G}$, the critical momentum $\sim 1/\theta_{\star}$~GeV is in the non-relativistic regime. Even if at lower energies the particle distribution is hard, since pressure is always dominated by relativistic particles, one should not expect a strong back-reaction of accelerated particles over the fluid inside the superbubble, compared to the case where collective acceleration effects are not taken into account. However, if the magnetic field pressure is close to equipartition with the thermal pressure as suggested by \cite{Parizot2004a}, and provided that the turbulence index~$q$ is sufficiently low, then the impact of particles on their environment has to be investigated. More generally, if $q$ is low enough and/or $B$ is high enough, then the superbubble can no longer be regarded as a sum of isolated supernovae, but acts as a global accelerator, producing hard spectra over a wide range of momenta.

One can wonder how solid these results are, given all the uncertainties in the data. In particular, the parameter $\theta_{\star}$ is very sensitive to the accelerator size $x_{\rm max}$. However $x_{\rm max}$ cannot be much lower than a few tens of parsecs (the typical size of the OB association) and cannot be much larger than 100~pc (the typical size of the superbubble). The maximal scale of the turbulence, $\lambda_{\rm max}$, is even more difficult to estimate, but it also ranges between those extrema. Determining precisely these spatial scales is complicated by the difficulty of estimating the supershell associated with a given cluster, all the more so since multiple bursts episodes have occurred (as is likely the case in 30~Doradus). In addition, $\theta_{\star}$~is directly proportional to the density, which is not always measured with good accuracy, but can usually be rather well constrained to within one order of magnitude. The upper and lower values of $\theta_*$ given in Table~1 reflect the uncertainties in these three key parameters. In the end, we believe that the results presented in Table~1 provide a good indication of whether or not collective effects will dominate inside the superbubble. Across the range of possible values of size and density, the main uncertainty in the critical parameter $\theta_{\star}$ is clearly due to our poor knowledge of the magnetic field (how strong the field is, how turbulent it is). It can be seen from Table~1 that for a given prescription of the magnetic turbulence, the values obtained for both Galactic and LMC clusters are not very different from one another. 
~\\


\section{Limitations and possible extensions}
\label{sec:limitations}


\subsection{Regarding shock acceleration physics}
\label{sec:limitations_shocks}

The potentially greatest limitation of our model is its use of a linear model
for regular acceleration: we have not considered the back-reaction of accelerated particles
on their accelerator, whereas cosmic rays may easily modify the supernova remnant shock and
therefore the way in which they themselves are accelerated \citep{Malkov2001c}.
Since non-linear acceleration is a difficult problem, only a few models are available,
such as the time-asymptotic semi-analytical models of \citet{Berezhko1999a} or \citet{Blasi2005b},
and the time-dependent numerical simulations of \citet{Kang2007a} or \citet{Ferrand2008a}.
We will include one of these non-linear approaches in our Monte Carlo framework
in extending our current work.
We can already note that non-linear effects tend to produce concave spectra,
softer at low energies and harder at high energies than
the canonical power-law spectrum, and may thus compete with reacceleration
and escape effects that we have shown to have opposite effects.
Moreover, non-linearity also occurs regarding the turbulent magnetic field 
(mandatory for Fermi process to scatter off particles),
which remarkably can be produced by energetic particles themselves 
by various instabilities. This difficult and still quite poorly understood process
has been studied by means of MHD simulations (\citealt{Jones2006a}), 
semi-analytical models (\citealt{Amato2006a}),
and Monte Carlo simulations (\citealt{Vladimirov2006a}).

Another limitation is that only strong primary supernova shocks have been
considered (of compression ratio $r=4$), but since superbubbles are very
clumpy and turbulent media, many weak secondary shocks are also expected
(of $r<4$). The compression ratio $r$ depends on the Mach number
$M_{S}$ according to
\begin{equation}
r=\frac{4\: M_{S}^{2}}{M_{S}^{2}+3}\:,
\label{eq:r(Ms)}
\end{equation}
where
\begin{equation}
M_{S}=\frac{v_{S}}{c_{S}}\simeq50\left(\frac{v_{S}}{5000\:\mathrm{km/s}}\right)\left(\frac{T}{10^{6}\:\mathrm{K}}\right)^{-1/2}
\label{eq:Ms}
\end{equation}
and $u_{S}$ is the shock velocity (of many thousands of km/s in the early stages of a remnant evolution) 
and $c_{S}$ is the speed of sound in the unperturbed upstream medium
(as high as a few hundreds of km/s in a superbubble because of the high temperature
$T$ of a few millions of Kelvin). In the linear regime, the slope of accelerated particles is
determined solely by $r$ according to Eq.~(\ref{eq:s1}). 
In superbubbles, 
primary supernova shocks have $M_{S}\simeq50$ and already $r\simeq4$, leading to $s\simeq4$; 
but a secondary shock of say $M_{S}\simeq5$ has only $r\simeq3$, leading to $s\simeq4.5$. 
We note that although weaker shocks produce softer individual spectra, 
being more numerous they may help to produce hard spectra by repeated acceleration,
so that their net effect is not obvious. To begin their investigation,
we added a weak shock at each time-step immediately following
a supernova (except if another supernova occurs at that moment), 
of compression ratio randomly chosen between 1.5 and 3.5. 
For regular acceleration alone, important differences are seen 
between simulations including only strong shocks, or only weak shocks, or both. 
But once combined with stochastic acceleration and escape, 
these differences are no longer evident. 
We have repeated our 720 simulations at medium resolution and 
observed that our two indicators (momentum of transition and minimal slope) 
remain globally unchanged. 
The shape of cosmic-ray spectra thus seems to be mostly determined 
by the interplay between reacceleration and escape, acceleration at shock fronts 
acting mostly as an injector of energetic particles. 
We note that, before supernova explosions, the winds of massive stars, 
not explicitly considered in this study, may also act as injectors in the same way,
as they have roughly the same mechanical power integrated over the star lifetime.

Finally, one may question our particular choice of stellar evolution models,
but we believe that possible variations in the exact lifetime of massive stars
would bring only higher order corrections to the general picture that we have obtained.
We also note that we have implicitly considered that stars are born at the same time,
and then evolve independently, while in reality star formation may occur through 
successive bursts within a same molecular cloud, which could be sequentially
triggered by the first explosions of supernovae. 
Another possible amendment to our model is  
that stars of mass greater than 40~solar masses may end their life without collapsing, 
and thus without launching a shock. We have repeated our 720 simulations at low resolution 
considering the occurrence of supernovae only for $m<40\: m_{\odot}$,
and checked that our two indicators remain globally unchanged. 
This seems consistent with the shape of the IMF (there are very few stars of very high mass)
and the shape of star lifetimes (stars of very high mass have roughly the same lifetime).


\subsection{Regarding inter-shock physics}
\label{sec:limitations_inter}

We use an approximate model of stochastic acceleration,
because of the use of relativistic formulae and the neglect of energy losses, 
to be able to use results from \citet{Becker2006a}. 
However, we note that, in terms of stochastic acceleration, the relativistic regime
is reached when $m_{p}v\gg m_{p}v_A$, where $v$ is the particle velocity and $v_A$ the Alfv{\'e}n velocity
\begin{equation}
v_{A}=\frac{B}{\sqrt{\mu_{0}\rho}}\simeq2.10^{7}\:\mathrm{cm.s^{-1}}\:\left(\frac{B}{10\:\mathrm{\mu G}}\right)\left(\frac{n}{10^{-2}\:\mathrm{cm^{-3}}}\right)^{-\frac{1}{2}}\:,
\label{eq:vA}
\end{equation}
and in a superbubble this condition is met for $p\gg1\:\mathrm{MeV}$, since $v_{A}/c\simeq10^{-3}$. 
Although we could of course implement more involved models of transport,
we emphasize that our main objective was to find the key dependences of the problem, 
and we have shown that it is mainly controlled by the parameter~$\theta^{\star}$.
Regarding losses, the formalism of \citet{Becker2006a} allows for systematic losses, 
but for mathematical convenience these are supposed to occur at a rate~$\propto p^{q-1}$,
which can describe Coulomb losses only in the very special case of $q=2$.
But proton losses above 1~GeV are dominated by nuclear interactions \citep{Aharonian1996a} 
with a typical lifetime of $6.10^{7}\:\mathrm{yr}\:/n$, where $n$ is the density in $\mathrm{cm^{-3}}$,
which is far longer than the superbubble lifetime given the low density $n\leq10^{-2}\:\mathrm{cm^{-3}}$
(but this might become a concern when cosmic rays reach 
the parent molecular clouds where $n>10^{2}\:\mathrm{cm^{-3}}$). 
At very low energies (around the MeV), ionization losses might also
be important and compete with stochastic reacceleration.

Finally, we note that most parameters 
are time-dependent, and might become considerably different at late stages.
For completeness, we have performed our simulations until the explosion of the longest lived stars, 
but over tens of millions of years the overall morphology and properties of the superbubble 
might change substantially as it interacts with its environment.
As long as the typical evolution timescale of relevant parameters is longer than our time-step
$dt=10\:000\:\mathrm{yr}$, their variation can be taken into account
simply by varying the value of $\theta^{\star}$ accordingly. 
Otherwise, direct time-dependent numerical simulations similar to those of \citet{Ferrand2008a}
will be necessary.


\section{Conclusions}
\label{sec:conclusions}

Our main conclusions are as follows:
\begin{enumerate}
\item Cosmic-ray spectra inside superbubbles are highly variable: 
at a given time they depend on the particular history of a given cluster.
\item Nevertheless, spectra follow a distinctive overall trend, produced by
a competition between (re-)acceleration by regular and stochastic
Fermi processes and escape: they are harder at lower energies ($s<4$) and
softer at higher energies ($s>4$), shapes that are in agreement 
with the results of~\citet{Bykov2001b} based on different assumptions\footnote{
\cite{Bykov2001b} considers acceleration of particles by large-scale motions of the magnetized plasma inside 
the superbubble, which depends on the ratio $D_u/D_x$ where $D_x$ is the space diffusion coefficient, 
controlled by magnetic fluctuations at small scales, and $D_u=UL$ describes the effect of large scale turbulence, 
where $U$ is the average turbulent speed and $L$ is the average size between turbulence sources.}.
\item The momentum at which this spectral break occurs critically depends on the bubble parameters:
it increases when the magnetic field value and acceleration region size increase, 
and decreases when the density and the turbulence external scale increase,
all these effects being summarized by the single dimensionless parameter 
$\theta^{\star}$ defined by Eq.~(\ref{eq:Theta_def}).
\item For reasonable values of superbubble parameters, very hard spectra ($s<3$)
can be obtained over a wide range of energies, provided that superbubbles are highly magnetized 
and turbulent (which is a debated issue).
\end{enumerate}

These results have important implications for the chemistry inside superbubbles
and the high-energy emission from these objects.
For instance, in the superbubble Perseus~OB2 
there is observational evidence of intense spallation activity \citep{Knauth2000a} 
attributed to a high density of low-energy cosmic rays, 
but EGRET has not detected $\pi^{0}$-decay radiation, 
which places strong limits on the density of high-energy cosmic rays.
This is consistent with the shape of the spectra obtained in this work.
We are thus looking forward to seeing how new instruments such as Fermi and AGILE
will perform on extended sources such as massive star forming regions, 
which have recently been established as very high-energy sources.
In that respect, we make a final comment that the high intermittency of predicted spectra 
might explain the puzzling fact that some objects are detected while others remain unseen.


\begin{acknowledgements}
The authors would like to thank Isabelle Grenier and Thierry Montmerle for sharing their thoughts on the issues investigated here.
\end{acknowledgements}


\bibliographystyle{aa}
\bibliography{13520}

\end{document}